\documentclass[aps,preprint,nofootinbib,preprintnumbers,eqsecnum]{revtex4-1}
\pdfoutput=1

\usepackage{color}

\newcommand{\nb}[1]{\color{blue}}

\newcommand{\hl}[1]{\color{magenta}}

\usepackage[
	  pagebackref=false,
	  colorlinks=true,
      linkcolor=blue,
      urlcolor=blue,
      filecolor=black,
      citecolor=red,
      pdfstartview=FitV,
      pdftitle={},
        pdfauthor={Paolo Glorioso, Michael Crossley, Hong Liu},
        pdfsubject={},
        pdfkeywords={},
        pdfpagemode=None,
        bookmarksopen=true
      ]{hyperref}

\usepackage[normalem]{ulem}
\usepackage{amsmath}
\usepackage{enumerate}
\usepackage{amsfonts}
\usepackage{epsfig}
\usepackage{mathbbol}

\newcommand\half{{\ensuremath{\frac{1}{2}}}}
\newcommand\p{\ensuremath{\partial}}

\newcommand\vev[1]{{\ensuremath{\left\langle{#1}\right\rangle}}}

\newcommand{\be}{\begin{equation}}
\newcommand{\ee}{\end{equation}}
\newcommand{\bea}{\begin{eqnarray}}
\newcommand{\eea}{\end{eqnarray}}
\newcommand{\bega}{\begin{gather}}
\newcommand{\eega}{\end{gather}}

\newcommand{\bi}{\begin{itemize}}
\newcommand{\ei}{\end{itemize}}
\newcommand{\ben}{\begin{enumerate}}
\newcommand{\een}{\end{enumerate}}
\newcommand{\bca}{\begin{cases}}
\newcommand{\eca}{\end{cases}}
\newcommand{\bln}{\begin{align}}
\newcommand{\eln}{\end{align}}
\newcommand{\bst}{\begin{split}}
\newcommand{\est}{\end{split}}
\def\ie{\begin{equation}\begin{aligned}}
\def\fe{\end{aligned}\end{equation}}
\newcommand{\bma}{\le(\begin{matrix}}
\newcommand{\ema}{\end{matrix}\ri)}

\def\b{{\beta}}
\newcommand\ep{\epsilon}
\newcommand\sig{\sigma}

\newcommand\lam{\lambda}

\newcommand\om{\omega}

\newcommand\ga{{\ensuremath{{\gamma}}}}

\newcommand\de{{\ensuremath{{\delta}}}}
\newcommand\De{{\ensuremath{{\Delta}}}}

\newcommand\ka{\kappa}

\def\th{{\theta}}

\newcommand\Lra{{\Longrightarrow}}
\newcommand\ov{\over}
\newcommand\ha{{\half}}

\def\le{\left}
\def\ri{\right}

\newcommand\sD{{\ensuremath{{\mathcal D}}}}
\newcommand\sE{{\ensuremath{{\mathcal E}}}}

\newcommand\sG{{\ensuremath{{\mathcal G}}}}

\newcommand\sL{{\ensuremath{{\mathcal L}}}}

\newcommand\sN{{\ensuremath{{\mathcal N}}}}

\newcommand\sP{{\ensuremath{{\mathcal P}}}}

\newcommand\sJ{{\mathcal J}}

\newcommand\sT{{\mathcal T}}

\newcommand\vx{{\vec x}}

\newcommand{\fre}{{\mathfrak{e}}}

\newcommand{\fs}{{\mathfrak s}}
\newcommand{\ft}{{\mathfrak t}}

\newcommand{\Sch}{{\rm Sch}}

\begin{document}

\title{A quantum hydrodynamical description for scrambling and  many-body chaos}

\preprint{MIT-CTP/4975}

\author{Mike Blake, Hyunseok Lee, and Hong Liu}
\affiliation{Center for Theoretical Physics, \\
Massachusetts
Institute of Technology,
Cambridge, MA 02139 }

\begin{abstract}

\noindent Recent studies of out-of-time ordered thermal correlation functions (OTOC) in holographic systems and
in solvable models such as the Sachdev-Ye-Kitaev (SYK) model have yielded new insights into manifestations of many-body chaos.
So far the chaotic behavior has been obtained through explicit calculations in specific models.
In this paper we propose a unified description of the exponential growth and ballistic butterfly spreading of OTOCs 
across different systems using a newly formulated ``quantum hydrodynamics,'' which is valid at finite $\hbar$ and to all orders in derivatives.  
The scrambling of a generic few-body operator in a chaotic system 
is described as building up a ``hydrodynamic cloud,'' and the exponential growth of the cloud arises from a shift symmetry of the hydrodynamic action.
The shift symmetry also shields correlation functions of the energy density and flux, and time ordered correlation functions of generic operators from exponential growth, while leads to chaotic behavior in OTOCs. 
The theory also predicts an interesting phenomenon of the skipping of a pole at special values of 
complex frequency and momentum in two-point functions of energy density and flux. This pole-skipping phenomenon may be considered as a ``smoking gun'' for 
the hydrodynamic origin of the chaotic mode. We also discuss the possibility that such a hydrodynamic description could be a hallmark of maximally chaotic systems.

\end{abstract}

\today

\maketitle

\tableofcontents

\section{Introduction}

Chaotic phenomena are ubiquitous in nature. While much has been learned about 
chaos at a classical level,
%and chaos behavior provides a foundation for various hypothesis of classical statistical physics. 
its characterizations and manifestations at a quantum level are far less understood, % We expect that there is still much to be learned, 
especially in many-body systems.  

One characterization of chaos in quantum many-body systems is the Eigenstate Thermalization Hypothesis (ETH)~\cite{Deutsch,Srednicki,Srednicki1999,Rigol,Review} 
which says that for a chaotic system, highly excited energy eigenstates should behave like a thermal state. ETH is a powerful statement, implying that chaos underlies thermodynamical behavior of an isolated quantum statistical system.
But it is not easy to work with or to check explicitly in practice as energy eigenstates of many-body systems are hard to come by.\footnote{%See recent studies in CFTs~\cite{} where analytic results can be obtained. 
See also~\cite{Dymarsky:2017zoc} for a recent proposal for another manifestation of quantum chaos in many-body systems.} 
More recently, studies of out-of-time ordered thermal correlation functions in holographic systems and
in solvable models like Sachdev-Ye-Kitaev (SYK) model have yielded new insights into manifestations of many-body chaos~\cite{kitaev0,Roberts, Shenker1,Shenker2,Polchinski:2016xgd,MSS, MSY, MS, Jensen,Engelsoy:2016xyb,Kitaev,Jevicki:2016bwu,Jevicki:2016ito}. 

More explicitly, consider a quantum many-body system at a finite temperature $T_0 ={1 \ov \beta_0}$, and 
the expectation value of the square of  the commutator of two generic few-body operators $V$ and $W$ %which behaves as 
\bega \label{iem}
C(t) =  - \vev{[V (t), W(0)]^2} = C_{1} (t) - C_{2} (t)   \\ 
\label{TOC}
C_1 (t) = \vev{V(t) W(0) W(0) V(t)} + \vev{W(0) V(t) V(t) W(0)}\\
C_2 (t) =  \vev{V(t) W(0) V(t) W(0)}
+ \vev{W(0) V(t) W(0) V(t)}
\label{OTOC}
 % \cr
%&=& \vev{V(t) W(0) W(0) V(t)} + \vev{W(0) V(t) V(t) W(0)} - \vev{V(t) W(0) V(t) W(0)}
%- \vev{W(0) V(t) W(0) V(t)}
%\sim e^{\lam (t- {x \ov v_B})} 
\end{gather}
where $C_1$ and $C_2$ are respectively referred to as the time ordered (TOC) and out-of-time ordered (OTOC)  
correlation functions.

\begin{figure}
\begin{center}
\resizebox{120mm}{!}{\includegraphics{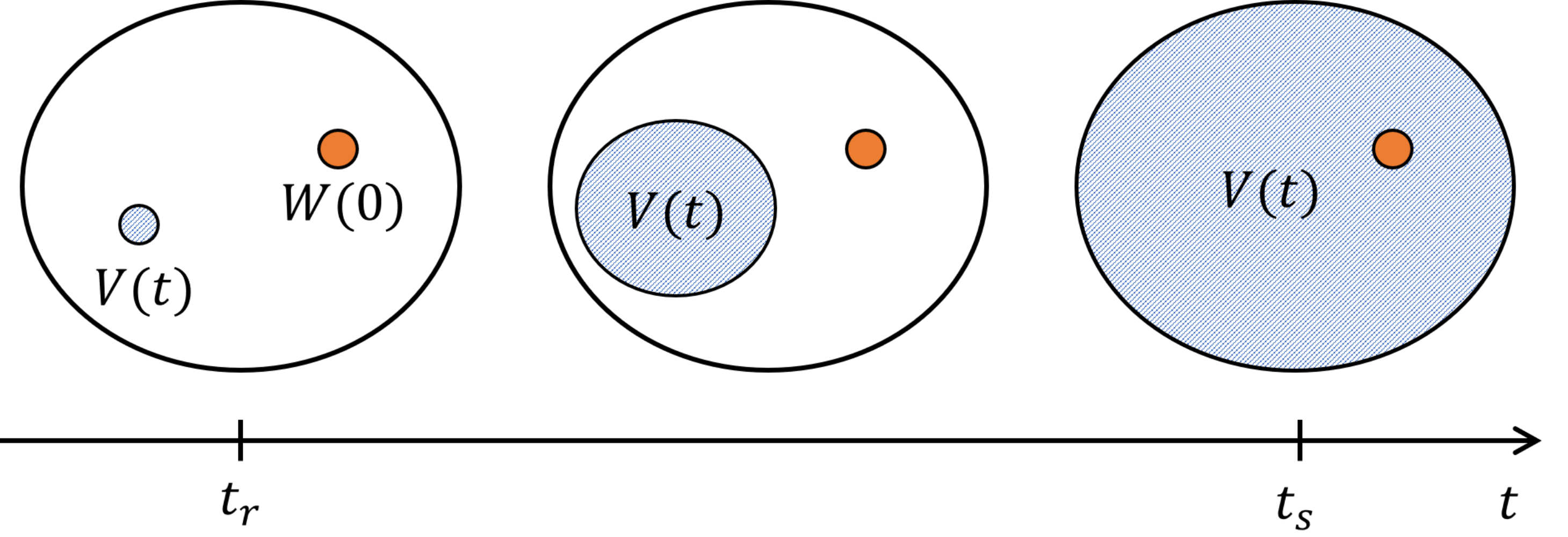}}
\caption{A cartoon of scrambling. The operator $V(t)$ expands in the space of degrees of freedom. $t_r$ is relaxation time, and $t_s$ is the scrambling time 
when $V(t)$ is essentially scrambled among all degrees of freedom. 
% at a constant rate $\lambda \sim \beta^{-1}_0$. 
}
\label{fig:scrambling}
\end{center}
\end{figure}
 
For convenience throughout the paper we will assume that both $V$ and $W$ are normalized such that they can be considered as dimensionless. When $t$ is small, $C(t)$ should be very small,
\be 
C(t) \sim {1 \ov \sN}, \qquad t \lesssim t_r
\ee
where $\sN$ denotes the total number of degrees of the system and $t_r$ is the characteristic relaxation time of the thermal equilibrium. 
For a chaotic large $\sN$ theory one typically expects that $C(t)$ will grow exponentially with time
\be \label{iem0}
C(t) \sim {1 \ov \sN} e^{\lam t} \qquad t_r \ll t \ll t_s
\ee
until the so-called scrambling time $t_s \sim {1 \ov \lam} \log \sN$ when $C(t)$ becomes of $O(1)$. 
The Lyapunov exponent $\lam$ has been shown to be bounded in a generic system by~\cite{MSS} (below we will set $\hbar$ and $k_B$ to $1$ throughout the paper)
\be
\lam \leq \lam_{\rm max} = {2 \pi k_B \ov \beta_0 \hbar} \ .
\ee
Physically one can interpret the behavior~\eqref{iem0} as due to scrambling. At $t=0$ operator $V(0) = V$ 
involves only a few degrees of freedom. Under time evolution, $V(t)$ expands in the space of degrees of freedom, i.e. it 
gets scrambled among more and more degrees of freedom. $C(t)$ keeps increasing until $V(t)$ is scrambled essentially among all degrees of freedom, when $C(t)$ becomes $O(1)$ and saturates (See Fig.~\ref{fig:scrambling}). The exponential behavior~\eqref{iem0} reflects that 
the scrambling procceds at a steady rate with ${1 \ov C(t)} \p_t C(t) = \lam = {\rm const}$. 

When we separate $V$ and $W$ also spatially, then at large distances~\eqref{iem0} appears to generalize to
\be \label{iem1}
C(t, \vx) \equiv  - \vev{[V (t, \vx), W(0)]^2}  \sim {1 \ov \sN} e^{\lam (t - {|\vx| \ov v_B})} 
\ee
where $v_B$ in~\eqref{iem1} is often referred to as the butterfly velocity which characterizes the scrambling/expansion of an operator in space. Equation~\eqref{iem1} is observed in a SYK chain and essentially all holographic systems~\cite{Roberts,Shenker2,Gu}.  In other systems different forms of spatial propagation are often seen instead \cite{Swingle,Aleiner,Patel}. For instance in~\cite{Swingle,Aleiner} chaos is described by a diffusive spreading around the exponential growth\footnote{In~\cite{Gu,Shenker2}, the behavior below has been observed as transient behavior to~\eqref{iem1}.}
\be \label{mm0}
C(t, \vx) \sim {1 \ov \sN} e^{\lam t - {|\vx|^2 \ov D_{0} t}} \ .
\ee 
There have also been studies using random unitary circuits which find other types of butterfly spreading, but not a nonzero Lyapunov exponent~\cite{Nahum:2017yvy, Khemani:2017nda}.

The behaviors~\eqref{iem0}--\eqref{mm0} have been found in many model systems, often through complicated model 
specific calculations. A  unified understanding of how they emerge across different systems is still lacking. 
It is the purpose of this paper to propose such an effective description which does not depend on details of a specific system. %using ``quantum hydrodynamics'' whose precise meaning will be elaborated more below.  
We will be interested in those systems for which $\lam \sim {1 \ov \hbar}$, so that the chaotic behavior is 
intrinsically quantum, i.e. does not have a straightforward semi-classical limit.
 \begin{figure}
\begin{center}
\resizebox{90mm}{!}{\includegraphics{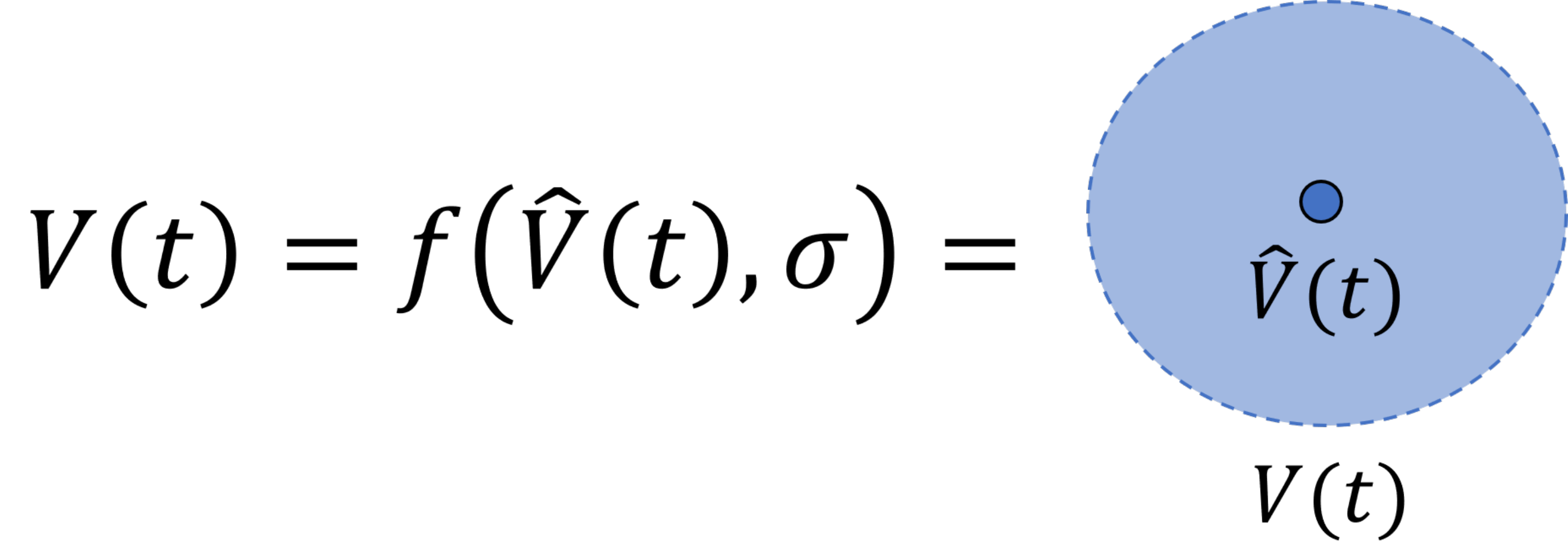}}
\caption{We propose that operator growth of $V(t)$ can be described in terms of a bare operator $\hat{V}(t)$ surrounded by an expanding hydrodynamic cloud.}
\label{fig:hydrocloud}
\end{center}
\end{figure}
\begin{figure}
\begin{center}
\resizebox{60mm}{!}{\includegraphics{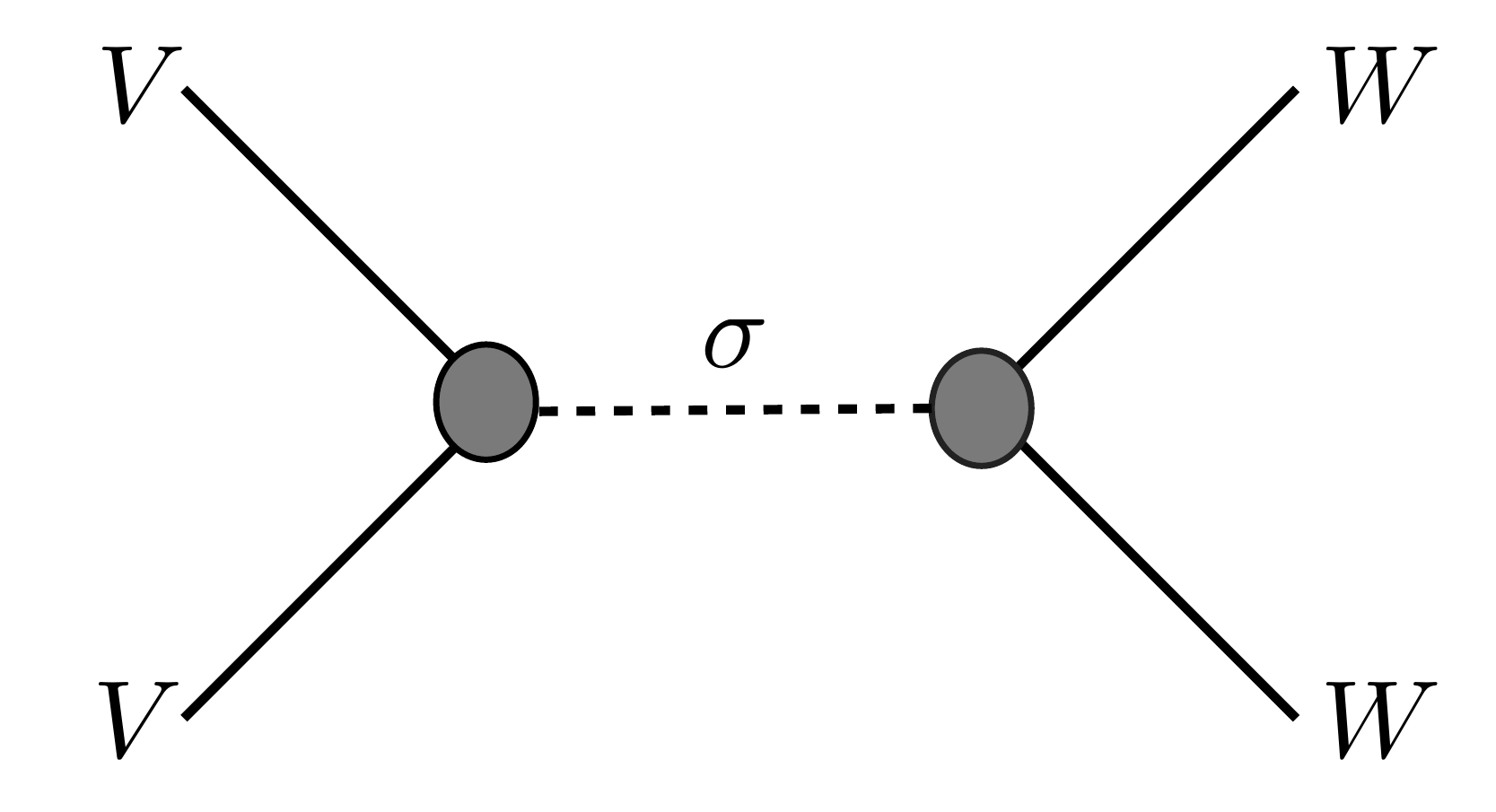}}
\resizebox{60mm}{!}{\includegraphics{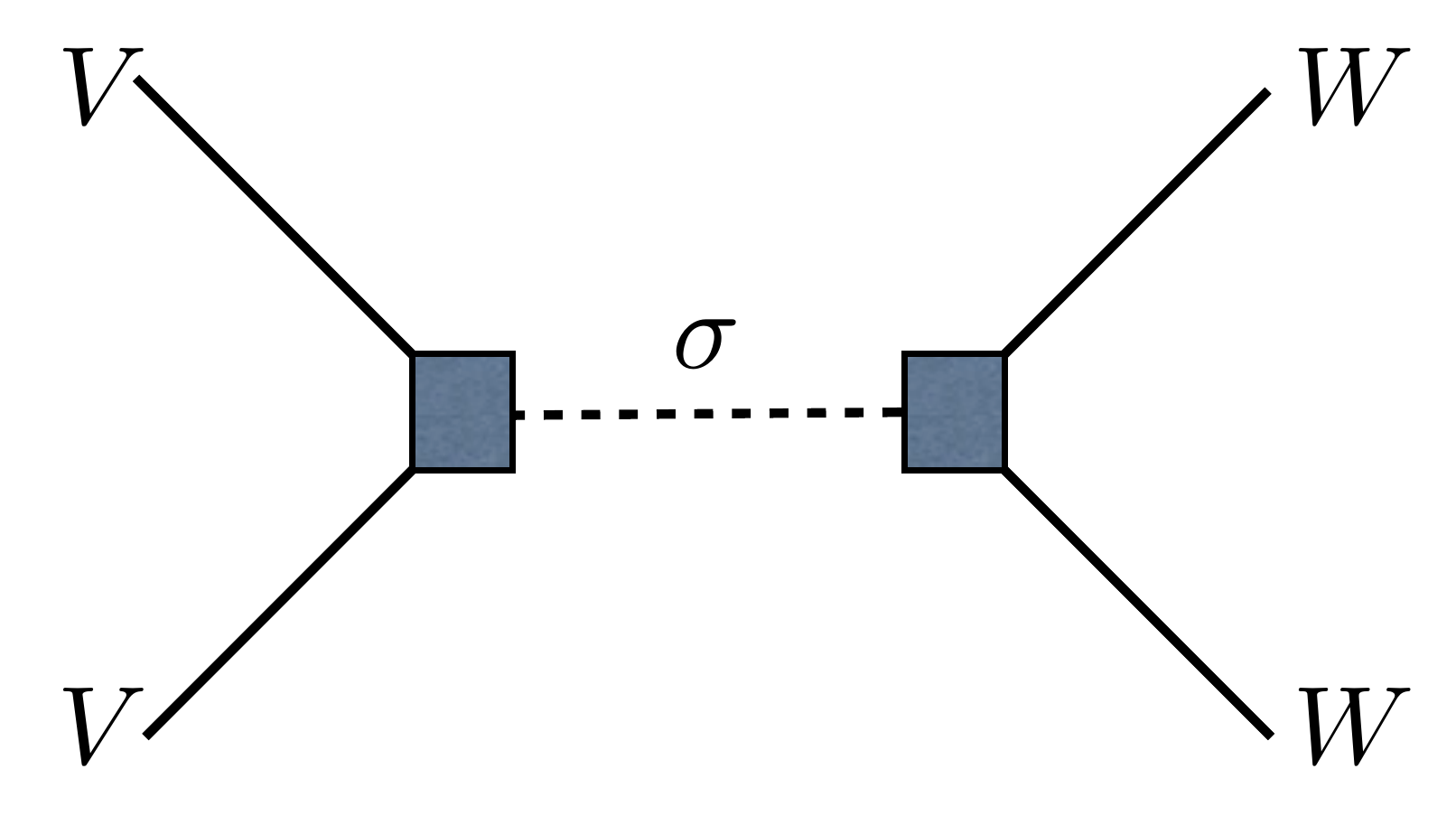}}
\caption{At leading order in large ${\cal N}$ correlation functions are controlled by exchange of hydrodynamic fields $\sigma(t)$. The only difference between time ordered (left) and out-of time-ordered (right) configurations is in the effective vertex describing the coupling to $\sigma(t)$.}
\label{fig:sigmainteraction}
\end{center}
\end{figure}
For such systems we propose the following effective description of chaos: 
\ben 
\item[A.]  To leading order in the limit $\sN \to \infty$,  the scrambling of a generic few-body operator in a chaotic system 
allows a coarse-grained description in which the growth of the operator can be understood as building up a ``cloud'' of some effective field $\sig$.  More explicitly, as indicated in Fig.~\ref{fig:hydrocloud}, $V (t)$ can be represented by a core operator $\hat V (t)$, which involves the degrees of freedom originally in $V$, dressed by a variable $\sig (t)$. % associated with energy conservation. 
\item[B.] The chaotic behavior~\eqref{iem0}--\eqref{mm0} of OTOCs can be understood from  exchanging and propagation of $\sig$ (see Fig.~\ref{fig:sigmainteraction}). 
%\item[C.] For maximal chaos $\lam = \lam_{\rm max}$, $\sig$ coincides with the hydrodynamics mode corresponding to {\it energy conservation.} 
\een
In this paper we will realise the above elements through developing a ``quantum hydrodynamic'' theory for chaos in which we identify $\sig$ with the hydrodynamic mode for energy conservation. As we will shortly discuss, this connection between the effective chaos mode and hydrodynamics can be motivated from the explicit calculations of OTOCs in holography and SYK models. Our general hydrodynamic theory not only provides a system-independent explanation of the chaotic behavior~\eqref{iem0}--\eqref{mm0} of OTOCs, but also leads to new predictions which can be explicitly checked. As we will {remark later} in the paper, likely the full content of this hydrodynamic theory only applies to systems {which are maximally (or nearly maximally) chaotic}. Nevertheless, we expect that various features associated with items A and B above may also apply to non-maximal chaotic systems. So throughout the paper we use a general Lyapunov exponent $\lam$ unless explicitly stated.

 %We will show that the Schwarzian is a special example of such a theory which happens to truncate at a finite order in derivatives. 

 There have been various hints for such an effective description, and possible connection with hydrodynamics.  In holographic systems, the scrambling and 
chaotic behaviors are realized through the backreaction of an in-falling particle to the spacetime geometry~\cite{Shenker1, Shenker2, Roberts} which can be interpreted in the boundary theory as building up a hydrodynamic cloud, see Fig.~\ref{fig:membrane}. 
In SYK, chaos is captured by a low energy effective action, the Schwarzian~\cite{kitaev0,Kitaev,MS,MSY,Jensen,Engelsoy:2016xyb,Jevicki:2016bwu,Jevicki:2016ito}. 
As already pointed out by K. Jensen~\cite{Jensen}, the Schwarzian can be considered as the effective action for a hydrodynamic mode associated with energy conservation (i.e. the $\sig$ variable mentioned above). In fact, Fig.~\ref{fig:hydrocloud} and Fig.~\ref{fig:sigmainteraction} have been implicit in various calculations performed using Schwarzian in~\cite{MSY} (see also~\cite{Kitaev}) for SYK. Here we propose that they should apply to general chaotic  systems.
Note the construction of the Schwarzian action from gravity~\cite{MSY} parallels that of construction of hydrodynamic action in~\cite{Crossley,Nickel:2010pr,deBoer:2015ija}. Furthermore,~\cite{Aleiner} extracted an unstable ``chaotic mode''  by developing a kinetic theory for OTOCs.
Finally in many holographic examples and non-holographic models the butterfly velocity $v_B$ in \eqref{iem1} and $D_0$ in~\eqref{mm0} appear to be related to the thermal/energy diffusion constant $D_E$ (for example see~\cite{MB,MB2,MB3,MB4,Davison,Gu, Patel, Swingle}).  Given that hydrodynamic degrees of freedom constitute a universal sector among all quantum many-body systems, it is then natural to search for a hydrodynamic origin of the chaotic behavior~\eqref{iem0}--\eqref{mm0} (see also~\cite{Lucas,Saso,Hartman} for other attempts to connect chaos and hydrodynamics).

Despite the above hints, at first sight there are nevertheless various important difficulties 
for a hydrodynamic description of chaotic behavior~\eqref{iem0}--\eqref{mm0}. We now discuss these difficulties and how to address them to help clarify and elaborate on our proposal:

 \begin{figure}
\begin{center}
\resizebox{50mm}{!}{\includegraphics{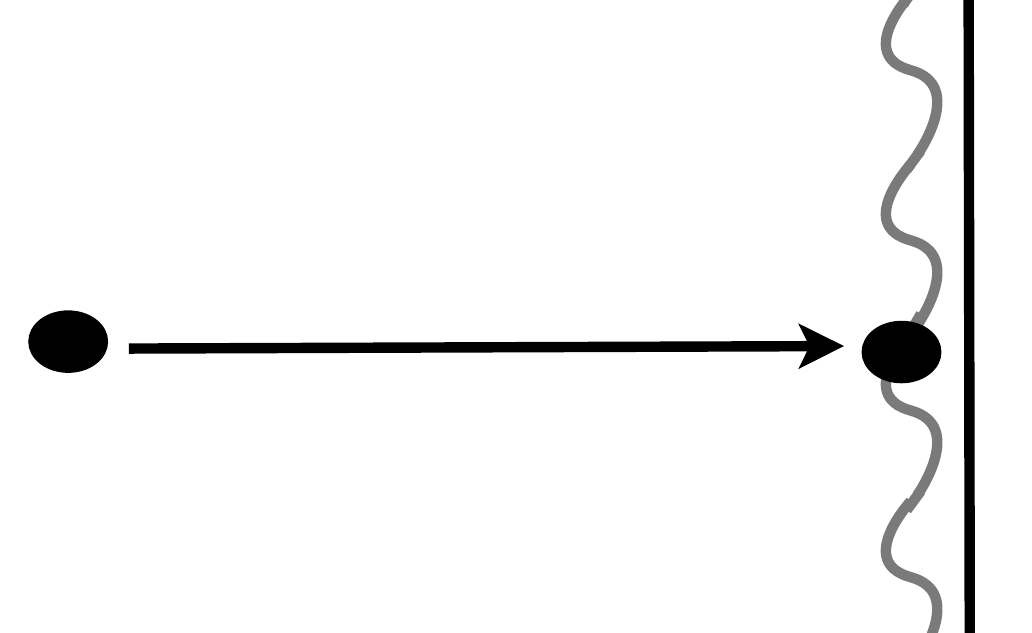}}
\caption{In holographic theories scrambling is described by the interaction of a particle with near-horizon degrees of freedom. The wavy line in the plot represents a stretched horizon while solid line the event horizon. As a particle falls into a black hole, from the perspective of an outside observer, the particle is ``dissolved'' into the thermal cloud of a stretched horizon. Such a process may be interpreted in the dual quantum field theory as building up a hydrodynamic cloud .}
\label{fig:membrane}
\end{center}
\end{figure}

\ben

\item Chaotic behavior lies outside the usual range of validity of hydrodynamics.
%The need for a quantum hydrodynamics to all derivative orders. 

Conventionally hydrodynamics is formulated as a low energy effective theory for gapless modes associated with conserved quantities,  valid for  time variation scales $\De t$ much larger than typical relaxation scales $t_r$, which for illustrative
purposes we will take to be of order $\b_0$ (e.g. in a strongly coupled system). 
It is written using a derivative expansion with expansion parameter ${\b_0 \ov \De t} \ll 1$. 
Such a description is inadequate for our purpose,\footnote{Of course in the standard hydrodynamic limit $\De t \ll \b_0$, 
the hydrodynamic equations  can exhibit chaotic behavior, such as turbulence. These infrared chaotic behavior has a smooth classical limit and has nothing to do with what we discuss in the paper which may be considered as ultraviolet chaos from hydrodynamics perspective.}
 as in~\eqref{iem0} the Lyapunov exponent $\lam$ is often of order ${1 \ov \b_0}$. Hence to capture the exponential growth~\eqref{iem0} one needs a formulation which is valid for 
$\De t \sim \b_0$. 
%In other words, one should formulate the theory to all derivative orders. 
Furthermore since we interested in quantum systems in which $\lam$ is proportional to $1/\hbar$, one needs a formulation which applies at quantum level with a finite $\hbar$, rather than the classical statistical limit which is normally taken. 

A quantum hydrodynamics which applies to the regime $\De t \sim \b_0$ %to all derivative orders 
can be obtained using the action formalism recently developed 
in~\cite{CGL,CGL1}.\footnote{See also~\cite{Bu:2014sia,Bu:2014ena} for discussions of extracting linearized hydrodynamic constitutive relations to all derivative orders from holography.}  The theory should be understood as obtained from integrating out all degrees of freedom of a quantum many-body system except for those associated with conserved quantities.  It can be nonlocal, but non-locality is only at scales of order $\b_0$.\footnote{For example it can incorporate quasi-normal modes.} This is due to the fact that for a generic system at a finite temperature, hydrodynamical modes are the only gapless degrees of freedom, thus the integrated out modes have energies or decay rates at least of order ${1 \ov \b_0}$. 
Working with such a theory is delicate as physics at scales of order $\b_0$
depends on microscopics of individual systems, which makes extracting universal information 
challenging. Nevertheless we will see that significant amount of universality can be obtained. 

In particular,  for the purpose of describing the chaotic behavior at leading order in $\sN \to \infty$ it is enough 
to work to quadratic order in perturbations from thermal equilibrium. 
Since the chaotic mode is only associated with energy conservation, to minimize technicalities in this paper we will consider a  theory whose only conserved quantity is energy conservation. Such a theory describes either a $(0+1)$-dimensional quantum mechanical system or a higher dimensional system with strong momentum dissipation at some microscopic scales. 
It should be straightforward, although technically more cumbersome, to write down a theory which has full energy-momentum conservation or other conserved quantities, which will be left for future work. 
 \begin{figure}
\begin{center}
\resizebox{70mm}{!}{\includegraphics{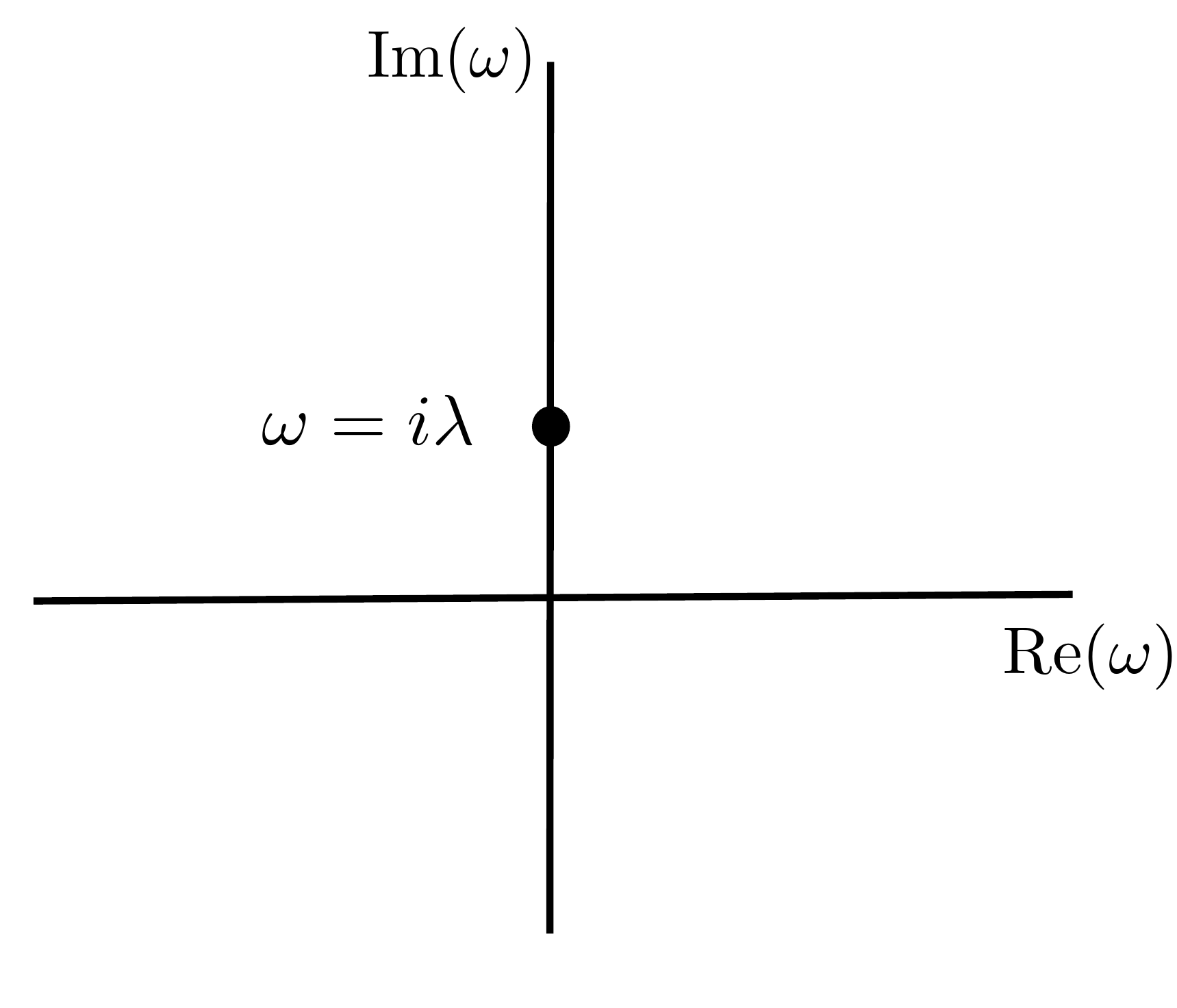}}
\caption{Chaos is described in our framework through a pole (at $\om = i \lam$) in the upper half complex frequency plane of the retarded Green function of hydrodynamic variable $\sigma(t)$. Such a pole does not indicate an instability, since it originates from a symmetry of the quantum hydrodynamics.}
\label{fig:chaospole}
\end{center}
\end{figure}

\item Can hydrodynamics of a stable system have an exponentially growing mode?

More explicitly, if exponential growth~\eqref{iem0} arises from exchanging and propagation of hydrodynamic variable $\sig$ 
associated with energy conservation, would that immediately lead to inconsistencies as the stress tensor 
%or other possible conserved currents 
certainly cannot have such exponential behavior? It turns out they can be perfectly consistent.

We propose that the quantum hydrodynamics for a chaotic system possesses a shift symmetry. On the one hand, this shift symmetry warrants that 
the retarded Green's function of $\sig$ has a pole in the upper half complex frequency plane for real momentum $k$ as indicated in Fig.~\ref{fig:chaospole}, which leads to exponentially growing behavior in real time. On the other hand, the structure of the quantum hydrodynamics is such that precisely the same symmetry ensures 
that such an exponential mode is invisible to the full stress tensor, and the retarded Green's functions of the stress tensor components have poles only in the lower half complex frequency plane for real $k$.

%\begin{figure}
%\begin{center}
%\resizebox{60mm}{!}{\include graphics{timeorderedinteraction.pdf}}
%\resizebox{60mm}{!}{\includegraphics{otimeorderedinteraction.pdf}}
%\caption{At leading order in large $N$ four point functions are dominated by exchange of hydrodynamic fields $\sigma(t)$. The only difference between time ordered (left) and out-of time-ordered (right) configurations is in the effective vertex describing the coupling.}
%\label{fig:sigmainteraction}
%\end{center}
%\end{figure}

\begin{figure}
\begin{center}
\includegraphics[scale=0.65]{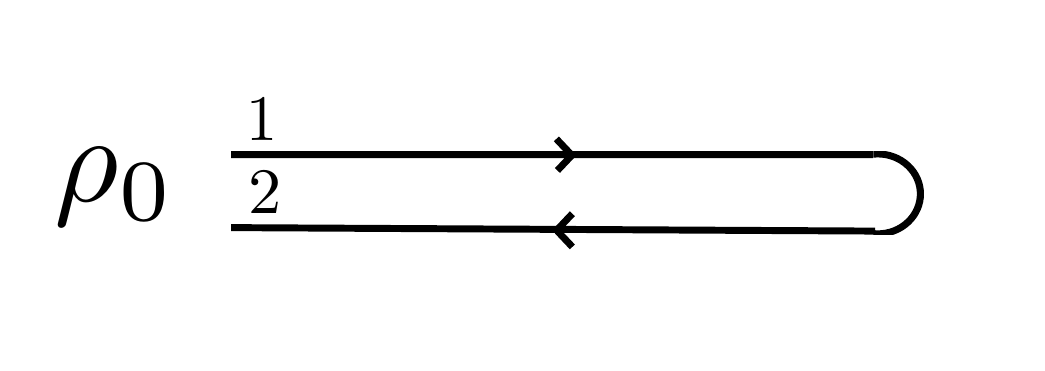} \quad
\includegraphics[scale=0.6]{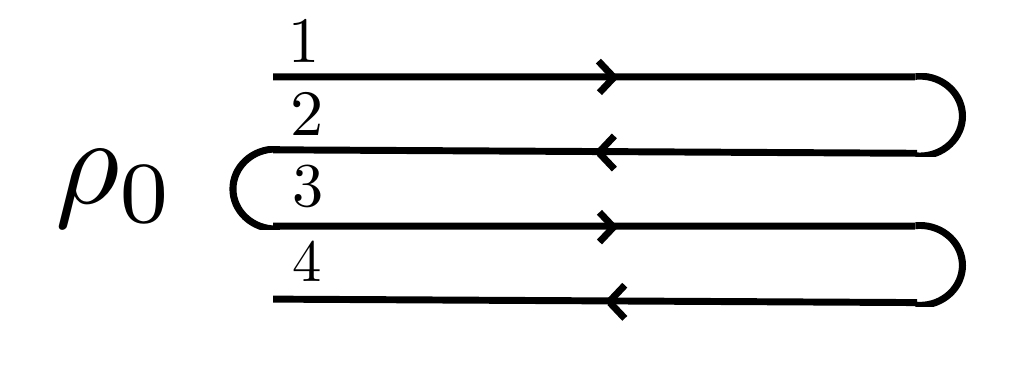}
%\resizebox{70mm}{!}{\includegraphics{skcontour.pdf}}
%\resizebox{70mm}{!}{\includegraphics{fourcontournumbered.pdf}}
\caption{Left: The hydrodynamic theory in a state $\rho_0$ is formulated directly in real time on the CTP contour. Right: OTOCs~\eqref{OTOC} require a contour of four segments.}
\label{fig:CTPcontour}
\end{center}
\end{figure}

\item In~\cite{CGL,CGL1} the hydrodynamics action is formulated as an effective theory for a system 
in some state $\rho_0$ defined on the closed time path (CTP), see Fig.~\ref{fig:CTPcontour}. 
How can a theory which is formulated on a closed time path, i.e. on a contour with two segments,
describe the chaotic behavior of OTOCs, which require a contour of four segments?

\begin{figure}
\begin{center}
\resizebox{140mm}{!}{\includegraphics{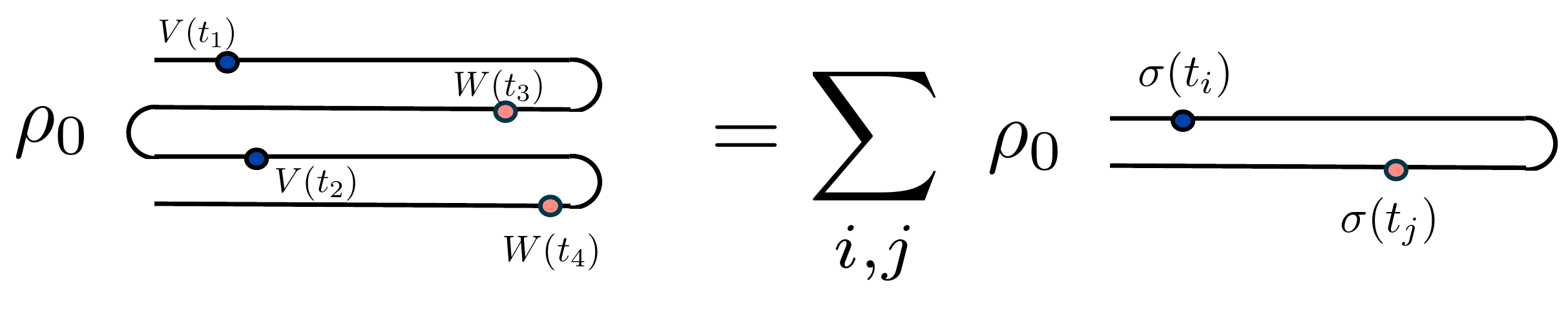}}
\caption{At leading order in large $\sN$ OTO four-point functions defined on a four-contour reduce to a sum of two-point functions of $\sigma(t)$. They can therefore be calculated from the effective action of the hydrodynamic field $\sigma$ on a CTP contour.}
\label{fig:otoreduction}
\end{center}
\end{figure}

The reason is simple. To leading order in $1/\sN$, only two-point functions of $\sig$ field are needed, as indicated in Fig.~\ref{fig:sigmainteraction}. 
Thus OTOCs on a 4-contour in the end reduce to a sum of two-point functions of $\sig$ on a two-contour which are in turn determined by the quantum hydrodynamics, see Fig.~\ref{fig:otoreduction}. In particular, the 
difference between the TOCs of~\eqref{TOC} and OTOCs of~\eqref{OTOC} lies in the 
precise structure of respective effective vertices, see Fig.~\ref{fig:sigmainteraction}. In TOCs, due to the shift symmetry in the coupling between the core 
operator $\hat V$ and its hydrodynamic dressing, the contributions 
of exponentially growing mode cancel while in OTOCs exponential mode survives and leads to~\eqref{iem0}--\eqref{mm0}.

\item Why does chaotic behavior have anything to do with energy conservation? 

Despite the aforementioned hints it is certainly not clear to what extent chaos should always be related to the hydrodynamic mode $\sigma(t)$ associated to energy conservation. For instance it is certainly the case that driven systems, in which energy conservation can be badly broken, can also be chaotic. {Chaos has also been studied using random unitary circuits~\cite{Nahum:2017yvy, Khemani:2017nda} which do not have energy conservation.}
  A possible explanation is that such a connection may be a feature of systems which are or are close to being maximally chaotic. We will elaborate more on this point in the conclusion section.

\een

%\item \textcolor{blue}{What predictions does a hydrodynamic description of chaos make?} 

\begin{figure}
\begin{center}
\resizebox{80mm}{!}{\includegraphics{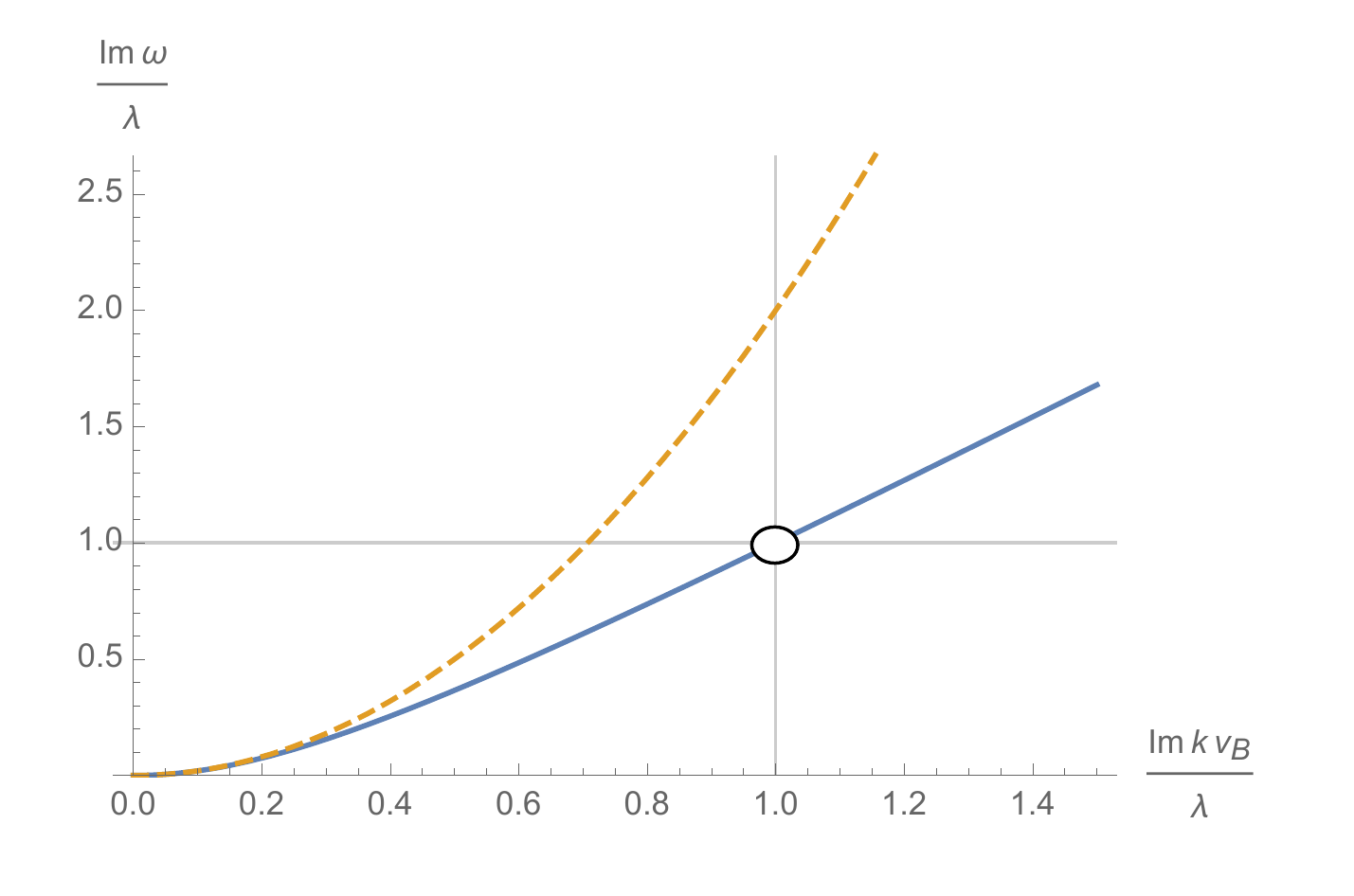}}
\caption{The hydrodynamic origin of chaos predicts  ``pole-skipping'' in the energy-energy correlation function:  following 
the line of poles which starts at small $\om, k$ as the energy diffusion pole along pure imaginary $k$ to a specific value of $k = i \lambda/v_B$ for which the pole would be at $\omega= i \lambda$, one finds the pole is not there!
In the figure the solid line denotes the line of poles, and the open dot indicates that at that particular point the pole is skipped.
The dashed line is the curve $\om = - i D_E k^2$ which coincides with the pole line for 
 small $\om, k$. 
}
\label{fig:vanishingpole}
\end{center}
\end{figure}

Now that we have motivated our hydrodynamic description of chaos we can describe the implications of such an effective theory. In particular, a key feature of this theory is that the mode $\sig$ which characterizes the energy conservation also describes the chaotic growth of general operators. This dual role leads to interesting predictions: many-body chaos also has bearings in two-point functions of energy density and energy flux, although in a subtle way. More explicitly, the theory predicts that:

\ben 
\item[a.]  The butterfly velocity $v_B$ is determined from the diffusion kernel. More explicitly, in a system with only energy conservation, one can introduce a general nonlocal diffusion kernel $\sD$ to relate energy flux $\sJ_i$ and the energy density $\sE$ as
\be \label{hnk}
\sJ_i = - \sD (\p_t, \p_i^2) \p_i \sE , \qquad \sD (\om, k^2)= D_E + O(\om, k^2)
\ee
where the leading order term in the small $\om, k$ expansion of $\sD$ is the energy diffusion constant $D_E$. 
One then finds that $v_B$ can be obtained from the solution to the following equation as
\be\label{vb}
\lam - k_C^2 \sD(i \lam, -k_C^2) = 0, \qquad v_B = {\lam \ov k_C} \ . 
\ee 
 %which is defined for general frequencies and momenta.
 In some special systems, $\sD$ collapses to a constant, i.e. higher order $\om, k^2$ terms in the second equation of~\eqref{hnk} all vanish. In such cases $v_B$ is then related to $D_E$ as $v_B^2 = D_E \lam$ that is seen in examples of SYK chains \cite{Gu:2017njx, Davison}. 
\item[b.] The pole line of energy density two-point function which originates from the diffusion pole $\om = - i D_E k^2$ (with $D_E$ the energy diffusion constant) for small $\om, k$ passes through the following point in the complex $\om$-$k$ plane  
\begin{equation}
\label{relationintro}
\omega(i k_C) = i \lambda , \qquad k = i k_C , \qquad  k_C \equiv  \frac{\lambda}{v_B}  \ .
\end{equation}
See Fig.~\ref{fig:vanishingpole}. 
\item[c.]  Precisely as the pole line passes through the specific point~\eqref{relationintro} the theory predicts that the residue of this pole should vanish. In other words, at that point, the pole is in fact skipped. This phenomenon, which we refer to as ``pole-skipping'', allows both $\lambda$ and $v_B$ to be directly calculated from knowledge of the energy-energy two point function, without the need to calculate OTOCs (see Fig.~\ref{fig:vanishingpole}). Such a behavior can be seen to happen in SYK chains and certain 
holographic systems~\cite{Gu,Saso}.
\een
The above predictions may be considered as a ``smoking gun'' for the hydrodynamic origin of the chaotic mode, and provide a simpler way than OTOCs to extract the Lyapunov exponent $\lam$ and butterfly velocity $v_B$.

The plan of the paper is as follows. In Sec.~\ref{sec:hydro} we introduce the quantum hydrodynamics for energy conservation. 
In Sec.~\ref{sec:cEFT} we introduce a shift symmetry to characterize chaotic systems. In Sec.~\ref{sec:coeft} we study correlation functions of the hydrodynamic variable in systems with a shift symmetry. In Sec.~\ref{sec:pole} we explain the phenomenon of pole skipping. In Sec.~\ref{sec:corre} we study TOCs and OTOCs of generic operators. %In Sec.~\ref{sec:max} we discuss some hints that the formalism developed here is most suitable for maximally chaotic systems.
We conclude in Sec.~\ref{sec:conc} with 
a summary and discussion. We have also included a few Appendices for various technical details.

\section{Quantum hydrodynamics for energy conservation} \label{sec:hydro} %Time re-parameterizations and 

Consider a quantum many-body system in a ``liquid'' phase for which the only gapless degrees of freedom are those associated with conserved quantities. The effective theory for these low energy degrees of freedom is hydrodynamics. 
In this section we discuss quantum hydrodynamics for a system whose only conserved quantity is energy, %and 
%its explicit quadratic action, 
following the general formulation of~\cite{CGL,CGL1}. %That is, we assume there is no other conserved quantity than energy. 
Such a theory applies to systems which have some way to strongly dissipate spatial momenta at some microscopic scales, but at macroscopic level still have spatial translation invariance.
With all spatial dependence dropped, the theory describes $(0+1)$-dimensional quantum mechanical systems. 
%It is enough to use this theory to illustrate all our essential points regarding scrambling and chaos.
Our discussion in this and subsequent sections can be readily generalized to systems with momentum conservations or other global symmetries.

\subsection{General quantum hydrodynamics formulation}

In~\cite{CGL,CGL1} the hydrodynamics action is formulated as an effective theory for a general statistical system 
in some state $\rho_0$ defined on the closed time path (CTP), see left plot of Fig.~\ref{fig:CTPcontour}. 
Let us first review the case with full energy-momentum conservation.

The formulation is reminiscent of the standard Lagrange description of fluid flows. One introduces a fluid spacetime with coordinates $\sig^A = (\sig^0, \sig^i)$ where $\sig^i$ can be interpreted labels of each fluid element and $\sig^0$ as the ``internal clock'' of a fluid element. The hydrodynamical degrees freedom are given by $X^\mu_{1,2} (\sig^A)$, which describe motions of fluid elements 
along two segments  of the CTP contour,\footnote{A single copy of such kind of variables was already used in~\cite{Herglotz}
and more recently in~\cite{Dubovsky:2005xd,Nickel:2010pr} for ideal fluids. The doubled copies for CTP contour were first used in~\cite{Grozdanov:2013dba} to describe dissipative effects.} and their equations of motion correspond to energy-momentum conservations 
associated with the two segments. One also introduces an inverse temperature $\b (\sig^A)$. 
Below we will denote the coordinates in physical spacetime as $x^\mu = (t, x^i)$.

{For the majority of this paper we will discuss systems with only energy conservation, in which case one can identify the spatial components of the fluid and physical spacetimes as $\sigma^i = x^i$. As we will explain, the hydrodynamic theory for such systems can be formulated in physical spacetime in terms of fluid time $\sigma^0(t,x^i)$ and the antisymmetric combination $X_a^0 = X^0_{1} - X^0_{2}$ which describes quantum statistical noises. We will ultimately drop the superscript $0$, and hence obtain a theory in which the fundamental dynamical field is the variable $\sigma = \sigma^0$ highlighted in our introduction. For completeness we first review various aspects of the general formulation of the hydrodynamic theory constructed in~\cite{CGL,CGL1}, before specializing to the simplified setting in which there is only energy conservation.}

To construct the hydrodynamic action consider turning on external metrics $g_{1\mu \nu}$ and $g_{2\mu \nu}$ associated with the two legs of the CTP. 
The hydrodynamic action $I_{\rm hydro}$ is required to depend on $X_{1,2}^\mu$ only through the induced metrics in the fluid spacetime,
\be 
h_{sAB} = g_{s \mu \nu} (X_s) {\p  X_s^\mu \ov \p \sig^A}  {\p  X_s^\nu \ov \p \sig^B} , \qquad s = 1,2
\ee
i.e.  $I_{\rm hydro} = I_{\rm hydro} [h_1, h_2, \b]$. It is further required to be invariant under the following fluid spacetime diffeomorphisms 
\bega \label{tdiff}
\sig^0 \to \sig'^0 (\sig^0, \sig^i), \qquad \sig^i \to \sig^i  , \\
\sig^0 \to \sig^0, \qquad \sig^i \to \sig'^i (\sig^i)  ,
\label{sdiff}
\end{gather}
and unitarity conditions 
\bega\label{CC1}
 I^*_{\rm hydro} [h_1, h_2, \b] = - I_{\rm hydro} [h_2, h_1 , \b] , \\
 \label{CC2}
 I_{\rm hydro} [h_1 = h_2, \b] = 0 , \\
 \label{CC3}
{\rm Im} \, I_{\rm hydro} \geq 0 \ . 
\end{gather}
Finally $I_{\rm hydro}$ is required to satisfy a $Z_2$ dynamical KMS symmetry which imposes local equilibrium as well as  microscopic time reversal symmetry\footnote{Here we will assume the Hamiltonian of the underlying system is invariant under $\sP\sT$.}. It has a simple form when we use~\eqref{tdiff} to set the local inverse temperature to be
\be\label{jnn}
\b = \b_0 E_r, \qquad E_r = \ha (\sqrt{-h_{100}} + \sqrt{-h_{200}})  \ 
\ee
with $\b_0$ some constant scale.\footnote{For $\rho_0$ given by a thermal density matrix, $\b_0$ is simply the background inverse temperature.} $I_{\rm hydro}$ is then required to be invariant under the following $Z_2$ transformation 
\bega
%\tilde X_1 (-\sig,-x^i) = -  X_1 (\sig + i \th, x^i)  + i \th , \quad 
%\tilde X_2 (-\sig, - x^i) = - X_2 (\sig - i \hat \th, x^i)  - i \hat \th ,  \\
\tilde h_{1AB} (-\sig^0, - \sig^i) =   h_{1AB} (\sig^0 + i \th, \sig^i) , \quad \tilde h_{2AB} (-\sig^0, - \sig^i) =   h_{2AB} (\sig^0 - i \hat \th, \sig^i),
\label{kms}
\end{gather}
for arbitrary $\th \in [0, \b_0]$ with $\hat \th = \b_0 - \th $. 

Various real-time correlation functions of the stress tensor of the full system can be computed using the theory of $I_{\rm hydro}$. For example, the retarded and symmetric two-point functions are obtained by  
\be \label{reg}
\sG^{\mu \nu, \lam \rho}_R (x^\mu) = i \vev{\hat T^{\mu \nu}_r (x) \hat T^{\lam \rho}_a (0)}_{\rm hydro} , \qquad
\sG^{\mu \nu, \lam \rho}_S (x^\mu) = \vev{\hat T^{\mu \nu}_r (x) \hat T^{\lam \rho}_r (0)}_{\rm hydro} 
\ee
where we have introduced the symmetric and antisymmetric pieces of the energy-momentum tensor
\bega 
\hat T^{\mu \nu}_r = \ha (\hat T_{1 \mu \nu} + \hat T_{2 \mu \nu}), \qquad 
\hat T^{\mu \nu}_a = \hat T_{1 \mu \nu} - \hat T_{2 \mu \nu}, \\
 \hat T^{\mu \nu}_1 (x) =  {2 \ov \sqrt{-g_1}} {\de I_{\rm hydro}  \ov \de g_{1\mu \nu} (x)} , \qquad
 \hat T^{\mu \nu}_2 (x)  = - {2 \ov \sqrt{-g_2}}  {\de I_{\rm hydro}  \ov \de g_{2\mu \nu} (x)}  
\ . 
\end{gather}
%

%Note that this formulation is fully quantum mechanical. 
%Furthermore equations of motion of $X^{\mu}_{1,2}$ are equivalent to the conservation of $\hat T_{1,2}^{\mu \nu}$ respectively. 

% and the local inverse temperature $\b (\sig^A(x^{\mu}))$. 

\subsection{Systems with only energy conservation} \label{sec:eng}

Now consider a system with only energy conservation, in which case we will set $X_{1,2}^i (\sig^0, \sig^i)= \sig^i$ and identify $ \sig^i=x^i$. The remaining dynamical variables are then 
$X^0_{1,2} (\sig^0, x^i)$, $\b(\sig^0, x^i)$,
%$ and denote 
%identify the spatial coordinates in physical and fluid spacetimes $\sig^i = x^i$ %The fluid spacetime coordinates are  thus $(\sig^0, x^i)$ and physical spacetime coordinates are $(t, x^i)$. 
%The  are then only $X^0_{1,2} (\sig^0, x^i)$, $\b(\sig^0, x^i)$, 
with equations of motion of $X_{1,2}^0$ equivalent to energy conservations. Below for notational simplicity we will drop superscripts $0$ on both $X^0$ and $\sig^0$. 
Since there is only energy conservation it is enough to turn on only the following external metric components 
\be \label{jj1}
ds^2 = g_{\mu \nu} dx^\mu dx^\nu = - e^2 \le(dt - w_i dx^i \ri)^2 + (dx^i)^2 , \quad g_{00} = - e^2 , \quad g_{0i} = e^2 w_i  
\ee
with the induced metric in the fluid spacetime 
\bega \label{indm}
ds^2= h_{AB} d \sig^A d \sig^B = - E^2 (d \sig - v_i dx^i)^2 + (dx^i)^2, \\
E = e \p_\sig X , \qquad v_i = {1 \ov \p_\sig X} \le(w_i - \p_i X \ri) \ . 
\end{gather}
%where 
%\be 
%E = e \p_\sig X , \qquad v_i = {1 \ov \p_\sig X} \le(w_i - \p_i X \ri) \ . 
%\ee
The energy density $\sE$ and energy flux $\sJ_i$ %associated with two copies of CTP contour
 can then be defined as 
\be \label{defej}
\sE = - \hat T^0_0 = - {1 \ov \sqrt{-g} } \le(e {\de I_{\rm hydro} \ov \de e} -w_i {\de  I_{\rm hydro}  \ov \de w_i} \ri) , \qquad \sJ^i =- \hat T_0^i =
 {1 \ov \sqrt{-g} }  {\de  I_{\rm hydro}  \ov \de w_i}  \ .
\ee
In~\eqref{jj1}--\eqref{defej} for notational simplicities we have suppressed indices $1,2$ labeling the two legs of the CTP;
it should be kept in mind there are two copies of them. The symmetry conditions~\eqref{tdiff}--\eqref{kms} are as before except that~\eqref{sdiff} reduces to rigid rotational symmetries (i.e. residual symmetries compatible with~\eqref{jj1} and~\eqref{indm}).

To obtain an action which is invariant under~\eqref{tdiff}
 it is convenient to define the following variables\footnote{The discussion below can also be obtained from that in Sec. V A of~\cite{CGL} by setting various quantities associated with spatial directions (and charged sector) to zero.} 
\be \label{oo1}
E_r = \ha (E_1 + E_2), \;\; E_a = \log (E_1 E_2^{-1}) , \;\;
V_{ai} = E_r (v_{1i} - v_{2i}) , \;\; V_{ri} = E_r v_{ri}, \;\; v_{ri} = \ha (v_{1i} + v_{2i})  \ .
\ee
Note that under~\eqref{tdiff}, $E_a, V_{ai}$ transform as a scalar. We also define covariant time and spatial derivatives
\be 
D_\sig \phi = {\p_\sig \phi \ov E_r}, \qquad d_i \phi = \p_i \phi + v_{ri} \p_\sig \phi 
\ee
which take a scalar $\phi$ under~\eqref{tdiff} to scalars. 
$E_r, V_{ri}$ do not transform as a scalar under~\eqref{tdiff}. They can be combined to form 
\be \label{o3}
%D_i E_r 
\fs_i = {1 \ov E_r} (\p_i E_r + \p_\sig V_{ri}) , \qquad \ft_{ij} = E_r (d_i v_{rj} - d_j v_{ri}) 
\ee
which transform as scalars under~\eqref{tdiff}. 
Note that $\fs_i$ and $\ft_{ij}$ are not completely independent as 
\be \label{olm}
D_\sig \ft_{ij} = d_i \fs_j - d_j \fs_i \ .
\ee
Thus the Lagrangian density can be constructed using $\b, \fs_i, \ft_{ij}, E_a, V_{ai}$ and their $D_\sig, d_i$ derivatives.
For convenience of imposing the $Z_2$ dynamical KMS symmetry we will again fix the local temperature through~\eqref{jnn}. 
Note that setting~\eqref{jnn} breaks~\eqref{tdiff} to a $\sig$-independent shift
\be  \label{shsy}
\sig \to \sig + c (x^i)\  .
\ee

\subsection{Physical spacetime formulation without sources} 

The above discussion is a bit formal. For the majority of our purposes in this paper it will be sufficient to work with the theory in the absence of external sources. That is, we set
\be \label{bac}
e_1 = e_2 =1, \qquad w_{1i} = w_{2i} = 0 , 
\ee
and write the action in physical spacetime. %In this case the hydrodynamic formulation is significantly simplified.

In particular, we now introduce $X = \ha (X_1 + X_2)$ and $X_a = X_1 - X_2$ where $X$ can be interpreted as describing physical motions while $X_a$  quantum statistical noises. To write the action in physical spacetime {we identify $X = t$} and then invert $X (\sig, x^i)$ to obtain $\sig (t, x^i)$.\footnote{Instead of $X$ we now use $t$ in the arguments of $\sig$ as it is now time coordinate of physical spacetime.}
 Thus the dynamical variables we will use to formulate our theory are now $\sig (t, x^i)$ and $X_a (t, x^i) \equiv X_a (\sig (t, x^i), x^i)$. 
 The various quantities described in~\eqref{oo1}--\eqref{o3} can be written in physical spacetime as  
\bega
E_r %= \p_\sig X 
= {1 \ov \p_t \sig}, \quad E_a = \p_t X_a + O(a^3), \quad V_{ai} = - \p_i X_a (t, x^i)  + O(a^3) , \quad \fs_i ,  \ft_{ij}= 
O(a^2) %, \quad = O(a^2)
%\\
% v_{ri} = \p_i \sig + O(a^2), \quad V_{ri} = - (\p_i X)_\sig + O(a^2),  \quad D_i E_r = O(a^2) , \quad \ft_{ij} = O(a^2)
\end{gather}
and 
\be 
D_\sig \phi (\sig (t, x^i), x^i) = \p_t \phi (t, x^i) , \qquad d_i \phi (\sig (t, x^i), x^i) = \p_i \phi (t, x^i) + O(a^2) 
\ee
where $O(a^2)$ denotes terms containing at least two factors of noise field $X_a$. 

{The most general Lagrangian for the fields $\sigma, X_a$ constructed from these quantities can be expanded to quadratic order in noise field $X_a$ as}
\be \label{unj}
\sL_{\rm hydro}[\sigma,X_a] =  - H \p_t X_a - G_i \p_i X_a +  {i \ov 2} M_1 (\p_t X_a)^2 + {i \ov 2} M_2 (\p_i X_a)^2  + O(a^3) % \p_i^2  X_a 
\ee
{where $H$, $G_i$ depend on $\sigma$ only through the local inverse temperature $\beta$ (recall~\eqref{jnn})}
\be \label{ombv}
\b = {\b_0 \ov \p_t \sig}
\ee
{and its ordinary spatial and time derivatives e.g. 
\begin{equation}
H = H(\beta, \partial_{t} \beta, \partial_{i}^2 \beta,...)  \ .
\end{equation}
Likewise $M_{1,2}$ should be understood as differential operators constructed out of $\b, \p_i, \p_t$ and acting on the first factor of $X_a$. In writing down equation~\eqref{unj} we have imposed~\eqref{CC1}--\eqref{CC2}.}
%Note that Equation~\eqref{unj} starts with linear order in $X_a$ due to~ 

The quantities $H$ and $G_i$ are respectively the dynamical part\footnote{This can be seen explicitly by turning on external sources as in Appendix~\ref{app:a}.} of the energy density and energy flux
\be 
\label{physical}
\sE_r = H + \cdots , \qquad \sJ^i_r = G_i + \cdots
\ee
 where $\cdots$ include contact terms and contributions from noises. The equations of motions that follow from \eqref{unj} then reduce to energy conservation\footnote{{Note that the equation of motion corresponding to $X$ always contains at least one factor of $X_a$ and thus we can consistently set $X_a =0$. In the absence of unphysical $a$-type sources, this is the only solution which satisfies the boundary condition $X_a (t=+\infty) =0$. Sometimes  $a$-type sources are turned on as a mathematical device for obtaining various types of correlation functions, in which case $X_a$ can have nonzero solutions. In this paper we will not need any $a$-type sources.
 %In addition to \eqref{enfl} there can be other solutions to the equations of motion in which the noise field $X_a$ is non-zero. Such solutions would be relevant if one turns on unphysical a-type sources (i.e. sources for the $X_a$ field) somewhere on the CTP contour.
 }}
\be \label{enfl}
\p_t H + \p_i G_i  = 0, \qquad X_a = 0 \ .
\ee
Clearly the equilibrium configuration 
\be \label{eui}
\sig= t, \qquad \b = \b_0, \qquad X_a = 0 
\ee
is always a solution to~\eqref{enfl}. 

{For $(0+1)$-dimensional quantum mechanical systems we simply set all the spatial derivatives to zero, i.e} 
\be \label{actn}
\sL_{\rm hydro}[\sigma, X_a] = - H \p_t X_a + {i \ov 2}  M_1 (\p_t X_a)^2 + O(a^3), %\qquad H = H (\b, \p_t \b, \p_t^2 \b , \cdots) \ 
\ee
{and \eqref{enfl} reduces to}
\be \label{enfl1}
\p_t H =0  , \quad X_a =0 \ .
\ee

\subsection{Near-equilibrium quadratic action} 

%\footnote{In Appendix~\ref{app:a} we consider the theory in the presence of external sources in order to compute the full expressions for the hydrodynamic stress tensor including contact terms.}

For our later purpose, let us now consider the quadratic action near equilibrium. Expanding around~\eqref{eui} we have
\be \label{uj}
\sig = t + \ep (t, x^i), \qquad \b = \b_0 + \de \b , \qquad \de \b =  \b_0 (1 - \p_t \ep) , \qquad X_a = -\ep_a (t, x^i)  
\ee
and at linear order in $\de \b$ we can write $H$ and $G_i$ as 
\be \label{omb}
H = f_1 \de \b = - \b_0 f_1  \p_t \ep , \qquad G_i = h_1 \p_i \de \b  = - \b_0 h_1 \p_i \p_t \ep
\ee
where $f_1, h_1$ are differential operators built from $\p_t, \p_i$. Since $X_a^2$ terms in~\eqref{unj} are already quadratic 
order, thus all $\b$-dependence in $M_1, M_2$ should be set to $\b_0$. With this understanding, the quadratic action can be written as 
\be \label{onbm}
\sL_{\rm hydro} =  \ep_a  K % (f_1  \p_t + h_1 \p_i^2 ) \p_t  
\ep - 
{i \ov 2} \ep_a M \ep_a , \quad K = \b_0 (f_1  \p_t + h_1 \p_i^2 ) \p_t, \quad 
M = M_1 \p_t^2 + M_2 \p_i^2
\ee
{and \eqref{enfl} becomes}
\be \label{qeom}
(f_1  \p_t + h_1 \p_i^2 ) \p_t  \ep =0, \qquad \ep_a = 0 \ .
\ee
Imposing dynamical KMS symmetry~\eqref{kms} requires that $f_1, h_1$ and $M_{1,2}$ satisfy (see Appendix~\ref{app:a})
\be\label{fdt1}
\b_0 (f_1 - f_1^*)= - 2 i \tanh{i \b_0 \p_t \ov 2} M_1 ,  \quad
 \b_0 (h_1 + h_1^*) \p_t  =- 2 i \tanh{i \b_0 \p_t \ov 2} M_2  \ 
\ee
where $f_1^*$ is the differential operator obtained from $f_1$ by integrations by parts, i.e. $f_1^* (\p_t, \p_i) = 
f_1 (- \p_t, - \p_i)$. Note by definition in~\eqref{onbm} $M_1 = M_1^*$ and $M_2 = M_2^*$. 

In the above discussion $f_1, h_1, M_1, M_2$ can be nonlocal at the scale $\b_0$. To make connections to conventional 
theory of hydrodynamics, let us consider~\eqref{qeom} to leading order in derivative expansions of these differential operators
\be \label{ing}
f_1 = - {c_0 \ov \b_0^2} + \cdots, \quad h_1 =  {\ka \ov \b_0^2} + \cdots , \quad M_1 = c_1 + \cdots, \quad M_2 = c_2 + \cdots
\ee 
where $c_{0,1,2}$ and $\ka$ are constants. From~\eqref{omb} we can identify $c_0$ as the specific heat while 
$\ka$ as the thermal conductivity.
Note that~\eqref{CC3} requires $c_1, c_2 \geq 0$ while~\eqref{fdt1} 
implies that 
\be 
 {\ka \ov \b_0^2}=  {c_2 \ov 2} \geq 0 \ .  
\ee
Equation~\eqref{qeom} then becomes diffusion equation 
\be \label{ing1}
(\p_t - D_E \p_i^2) \p_t \ep = 0
\ee
with energy diffusion constant 
\be \label{km}
D_E = {\ka \ov c_0},  %= {\ka_T \ov c_v}, \quad \quad c_v = {c_0  \b_0^2} , 
%\quad \ka_T = \ka \b_0^2 \  .
\ee
which is precisely the Einstein relation.

By including higher order terms in the derivative expansion, one could systematically incorporate corrections to~\eqref{ing}--\eqref{ing1} as in conventional hydrodynamics. However, as we emphasized in the introduction, the advantage of our approach is that it is possible to formulate the effective hydrodynamic theory non-perturbatively in derivatives. Such a framework is necessary in order to discuss the theory on scales $\Delta t \sim 1/\lambda$. In the next section we use this effective action $I_{\mathrm{hydro}}[\beta,X_{a}]$ to introduce our proposal for a hydrodynamic description of chaos.

To conclude this section let us briefly discuss the $\sN$ scaling of the hydrodynamic action. 
The variables $X, X_a$ are $O(1)$, while the coefficient of each term in the full nonlinear action should 
be of order $O(\sN)$. Thus in~\eqref{onbm}, $K, M, c_0, \ka \sim O(\sN)$ while $n$-point functions of $\ep, \ep_a$ scale 
as
\be \label{scad}
\vev{\ep^m \ep_a^n}_{\rm hydro} \sim \sN^{-{n+m \ov 2}} \ .
\ee

\section{Hydrodynamic theories for chaotic systems}  \label{sec:cEFT}

The quantum hydrodynamic theory introduced in last section applies to any quantum liquids at a finite temperature 
with energy as the only conserved quantity. In this section we show that imposing certain additional symmetry 
on the action gives rise to exponentially growing behavior of hydrodynamic variable $\sig$. 
In next section we will show such exponential behavior will not affect correlation functions of the energy density and flux,
while in Sec.~\ref{sec:corre} we show that it leads to chaotic behavior in OTOC. 

%In this section we will make a proposal for a description of chaos in terms of the hydrodynamic effective action. We define chaotic systems as systems whose OTC have exponential growth at small $t$. As we will show later this can be traced to that the hydrodynamic equations of motion have a solution which is exponentially growing with $t$. \textcolor{blue}{Does this require large N?}. Note that naively this is quite surprising, since it suggests that the chaotic mode can be identified by studying the action on a single Schwinger-Keldysh contour. In this section we will search for the class of action which has exponential growth. 

%We will first study some simple examples using the derivative expansion and then consider a general theory. 

%\subsection{A proposal for quantum chaotic system} 

\subsection{Shift symmetry in $0+1$-dimension}\label{sec:hyal}

To illustrate the general idea, let us first consider a quantum mechanical system with no spatial directions. 
We propose that a chaotic system with Lyapunov exponent $\lam$ can be 
described by an effective theory of the form~\eqref{actn} with an additional shift symmetry. 
{More explicitly, let us introduce $u(\sigma(t))$ through   
\be 
u = e^{-\lam \sig}, \qquad \sig = -{1 \ov \lam} \log u 
\ee
and require the Lagrangian $\sL_{\rm hydro}[\sigma, X_a]$ in~\eqref{actn} to be invariant under a shift symmetry
\be \label{oem1}
u \to u + a
\ee
for arbitrary constant $a$. }

Let us look at some immediate implications of this shift symmetry. Recall that equilibrium state is described by $\sig = t$, which 
in terms of $u$, is $u = u_0 = e^{-\lam t}$. From the shift symmetry, 
$u = e^{-\lam t} + a$ must also be a solution to equation of motion~\eqref{enfl1}. In terms of {$\b(\sigma(t))= \beta_0/\partial_{t} \sig$} this implies that 
\be \label{uie} 
\b = \b_0 + a \b_0 e^{\lam t} 
\ee
is a solution. At the linearized level, this corresponds to an exponentially growing solution in $\sigma(t)$
\be 
\sig = t - {a \ov \lam} e^{\lam t} + \cdots  
\ee
which we wish to propose as the origin of the chaotic behavior \eqref{iem0}.

To characterise hydrodynamic theories with this symmetry note that for $H$ to be invariant under the shift symmetry~\eqref{oem1} implies that it depends on {$u(\sigma)$} only through derivatives, i.e. 
\be \label{con1}
 H =  H (\p_t u, \p_t^2 u, \cdots)  \ .
 \ee
Now recall that by our original construction, $H$ depends on $\sig$ only through derivatives, i.e. by definition it is invariant under shift symmetry 
\be \label{oem2}
\sig \to \sig + c
\ee
 with $c$ a constant. Thus $H$ is characterized by two shift symmetries. Note in terms of $u$, the shift symmetry~\eqref{oem2} means scale invariance, i.e. 
\be \label{con2}
H (c u) = H (u)
\ee
for arbitrary constant $c$.

To have some intuition on the result action, let us expand $H$ in derivatives. The first few terms are  (primes below denote $t$ derivatives) 
\bea 
H &=& a_1 {u'' \ov u'} + {a_2 \ov 2} {u''^2 \ov u'^2} + a_3 \le({u'' \ov u'}\ri)'  + \cdots \cr
&=& \le(-{a_1 \lam \b_0 \ov \b} + {\lam^2 \b_0^2 a_2 \ov 2 \b^2} \ri)
+ \le(- {a_1 \ov \b} + \lam \b_0 {a_3 +  a_2 \ov \b^2} \ri) \b' + {\ha a_2 + a_3 \ov \b^2} \b'^2 - {a_3 \b'' \ov \b} 
\eea
where $a_{1,2,3}$ are constants. Note the term proportional to $\b'$ contains only one time derivative and is thus a friction-like dissipative term. 
It can be shown that it leads to entropy production. Setting $a_1 =0$ and $a_3 = -  a_2$, then $H$ is proportional to a Schwarzian 
\be \label{nd3}
H= - a_2 \Sch (u, t) =  a_2 \le( \ha {u''^2 \ov u'^2} - \le({u'' \ov u'}\ri)' \ri) %= \Sch (\sig, t) - {\lam^2 \ov 2} \sig'^2 
=a_2 \le( {\lam^2 \b_0^2 \ov 2 \b^2} - {\b'^2 \ov 2 \b^2} + {\b'' \ov \b}  \ri) \ 
\ee
which has a larger symmetry under $SL(2,R)$ transformations of $\sigma(t)$. When taking $\lam = \lam_{\rm max} =  {2 \pi \ov \b_0}$, this theory is reminiscent of the 
 Schwarzian action for SYK  and AdS$_2$ discussed in~\cite{kitaev0,MS,Jensen,MSY}. Indeed one can show that 
 the Lagrangian~\eqref{actn} with $H$ given by~\eqref{nd3} can be factorized into two copies of the Schwazian action,
see Appendix~\ref{app:b}.

To summarize, the shift symmetry~\eqref{oem1} warrants that the system has an exponentially increasing solution
of the form~\eqref{uie}. One immediate concern is that whether this will lead to instabilities. 
Note that $\sig(t)$, or equivalently the local inverse temperature $\beta(t)$, is not a physical
observable\footnote{{Although we refer to the field $\beta$ as a local inverse temperature this is merely an analogy with the usual formulation of hydrodynamics and does not imply $\beta$ can be directly measured. The precise meaning of $\beta$ is given by the mathematical identifications \eqref{jnn} and \eqref{ombv}. Physical quantities such as the energy density or energy flux are extracted using the relationships in \eqref{physical} between these quantities and $\beta$, which in general can be complicated}.}. It is a degree of freedom we use to parameterize non-equilibrium dynamics. Thus  exponentially behavior in $\sig$ does not have to imply instabilities. For example, despite the exponentially growing behavior in $\b$, the energy $H$ is always constant. In other words, the exponential behavior~\eqref{uie} carries no energy. Whether the exponential behavior shows 
up in other observables is a much trickier issue. 
In the context of Schwarzian for SYK and AdS$_2$, the $SL(2,R)$ symmetry is a global gauge symmetry, which 
ensures that the behavior~\eqref{uie} does not lead to any instability in any physical observables. 
Here we face similar issues; the exponentially growing behavior~\eqref{uie} should not lead to any instabilities 
in physical observables, but should lead to the exponential behavior of OTOCs~\eqref{OTOC} or commutators like~\eqref{iem}. 
These are nontrivial requirements, which we will discuss in detail in  Sec.~\ref{sec:corre}.

\subsection{Point-wise shift symmetry}

We now generalize the above discussion to systems with spatial dependence. 
The generalization is not unique. We first discuss a simplest possibility and then consider more general 
cases. 

%\subsubsection{Point-wise shift symmetry} 

As a simplest generalization of~\eqref{oem1}, we require the Lagrangian {$\sL_{\rm hydro}[\sig, X_a]$} to be invariant under an arbitrary spatial dependent shift symmetry  
\be \label{ene}
u(t,x^i) \rightarrow u(t,x^i) + a(x^i) \ . 
\ee
The physical interpretation of~\eqref{ene} is simple:  each fluid element has a separate shift symmetry. For such a Lagrangian it then follows that 
\be \label{exh}
%\sig = t - {\ep(x) \ov \lam} e^{\lam t} + \cdots \quad \to \quad
 \b = \b_0 + a(x^i) \b_0 e^{\lam t} 
 \ee
is a solution to equation of motion for arbitrary $a(x^i)$. Note that this statement holds at full nonlinear level.
 Invariance of~\eqref{unj} under~\eqref{ene} implies the invariance of 
 $H$ and $G_i$ under~\eqref{exh}, and thus the exponentially increasing behavior is invisible to both the energy density and energy flux.
 
With
\be 
\ga \equiv \p_t u
\ee
then invariance under~\eqref{ene} implies that 
\be \label{hbn}
H = H (\ga, \p_t \ga, \p_i \ga \cdots), \qquad G_i = G_i (\ga, \p_t \ga, \p_j \ga \cdots) \ .
\ee
The spatial dependent shift~\eqref{shsy} of $\sig$ now corresponds to $\ga \rightarrow c(x^i) \ga$
and thus both functions $H$ and $G_i$ in~\eqref{hbn} should be invariant arbitrary spatial dependent scaling of $\ga$, i.e. 
\be \label{hbn1}
H [\ga] = H [c(x^i) \ga], \qquad G_i [\ga] = G_i [c(x^i) \ga] \ .
\ee

As an illustration let us consider leading derivative expansion of $\sL_{\rm hydro}$ which satisfies~\eqref{hbn}--\eqref{hbn1} 
\begin{eqnarray}
{\cal L}_{\rm hydro} &=&  a_0 {\p_t \ga \ov \ga}  \partial_t X_a + a_1  \partial_{i} \le({\p_t \ga \ov \ga}\ri) \partial_{i} X_a  +O(a^2) \\
&=& - a_0 \bigg( \frac{\lambda \beta_0}{\beta} + \frac{\p_t \beta}{\beta}\bigg) \partial_{t} X_a - a_1 \partial_{i} \bigg( \frac{\lambda \beta_0}{\beta} + \frac{\p_t \beta}{\beta} \bigg) \partial_{i} X_a  +O(a^2)
\label{bnm}
\end{eqnarray}
%Note that the sign choice in the above equations are such that $D$ is positive (compare with~\eqref{diffu}). The energy density 
Comparing with~\eqref{unj} we thus find %and flux following from~\eqref{bnm} are  given by 
\be\label{ex1}
H = - a_0 \le(\frac{\lambda \beta_0}{\beta} + \frac{\p_t \beta}{\beta} \ri), \qquad G_i = - D \p_i H , \qquad D = - {a_1 \ov a_0}  \ .
\ee
Further expanding~\eqref{ex1} around equilibrium and comparing with~\eqref{omb} and~\eqref{km} we conclude that $D$ is precisely the thermal diffusion constant $D_E$. 

%We will see in later sections that point-wise shift symmetry leads to an interesting interplay between exponential growth and 
%diffusion which results in the ballistic propagation~\eqref{iem1}. 

\subsection{Effective action for a SYK chain}\label{sec:sykc}

As an application of the point-wise shift symmetry we now propose a real-time effective action for the 
SYK chains, as discussed for example in~\cite{Davison, Gu}. The effective action can be considered as a generalization of 
the $SL(2,R)$ invariant action~\eqref{nd3} to general dimensions, which incorporates diffusion. 
 %It is natural to expect that such an action should describe the spatial dynamics of 
 In the microscopic models of~\cite{Davison, Gu}, the theory has a large symmetry that corresponds to a separate $SL(2,R)$ transformation at each lattice point, which may be considered as enlarging~\eqref{ene} to point-wise $SL(2,R)$ symmetries. 
Such an action can be readily written down by taking
\be \label{ex2} 
H = -C \Sch(u,t) =  C\bigg( -{\b'^2 \ov 2 \b^2} + {\b'' \ov \b}  + {\lam^2 \b_0^2 \ov 2 \b^2}\bigg), \qquad G_i = - D \p_i H  
\ee 
where $C, D$ are some constants. 
The resulting action is invariant under $SL(2,R)$ transformations of the form
\be\label{sl2}
\mathrm{tanh} {\pi \sigma (t,x^i) \ov \b_0}  \rightarrow \mathrm{tanh} {\pi \tilde \sigma (t, x^i) \ov \b_0}= \frac{a (x^i)\, \mathrm{tanh} {\pi \sigma (t, x^i) \ov \b_0}+ b (x^i)}{c (x^i) \, \mathrm{tanh}{\pi \sigma \ov \b_0} + d (x^i)} , \qquad a d - b c = 1 \ .
\end{equation}
From~\eqref{uj} the quadratic action corresponding to~\eqref{ex2} near equilibrium can be written as 
\be \label{qw1}
\sL_{\rm hydro} =  C \ep_a \bigg( \lam^2 - \p_t^2  \bigg) (\partial_{t}  - D  \p_i^2  ) \p_t \ep
+ O(\ep_a^2) 
\ee
where $\ep^2_a$ term can be obtained from~\eqref{fdt1}. We see that $D$ again corresponds to energy diffusion constant $D_E$. 

It is interesting compare~\eqref{qw1} with the Euclidean quadratic action for SYK chains (which has $\lam = \lam_{\rm max}$) studied
in~\cite{Gu,Davison,Gu:2017njx}, which has the form
\begin{equation}\label{qw2} 
S_{\epsilon} = \tilde{C} \sum_{k, |\omega_n| \neq 0, {2 \pi \ov  \b_0}} |\omega_n| (|\omega_n| + D k^2)(\omega_n^2 - \lam_{\rm max}^2) |{\epsilon}(k, \omega_n)|^2  \ .
\end{equation}
Note the parallel between~\eqref{qw1} and~\eqref{qw2}. Also note that~\eqref{qw2} has one copy of $\ep$ while~\eqref{qw1} has two. While~\eqref{qw2} can be used to calculate Euclidean two-point functions in momentum space, it does not have a sensible 
continuation to Lorentzian signature.

To conclude this subsection, we note that in both examples~\eqref{ex1} and~\eqref{ex2} we have $G_i = - D \p_i H$. 
Of course in general this does not have to be the 
case. For example, combining~\eqref{ex1} and~\eqref{ex2} we can have 
\be \label{eno}
H = a_1 \frac{\partial_{t}^2 u}{\partial_{t} u} + a_2  \mathrm{Sch}(u,t) , \qquad
G_i = \p_i \le(c_1 \frac{\partial_{t}^2 u}{\partial_{t} u} + c_2 \mathrm{Sch}(u,t)\ri) \ 
\ee
for some constants $a_{1,2}$ and $c_{1,2}$.

\subsection{A general shift symmetry}

More generally we can require the action be invariant under 
\be \label{kn}
u  (t, x^i)  \to u  (t, x^i)  + f(t, x^i) 
\ee
for certain class of functions $f(t,x^i)$. Equation~\eqref{ene} corresponds to the class of $f$'s which are time independent, i.e. $f$ 
satisfies equation 
\be 
\p_t f = 0 \ .
\ee
Let us now suppose $f$ satisfies a more general differential equation 
\be \label{kn1}
\p_t f = \ka(\p_i) f
\ee
for some differential operator $\ka(\p_i)$ built from spatial derivatives. We require $\ka$ to contain at least one derivatives
such that a constant shift~\eqref{oem1} $f = c = {\rm const}$ is always allowed.
Writing $\sig = t + \ep (t, x^i)$ with $\ep $ infinitesimal, we then find the following $\ep$  
\be 
\ep  = -{f \ov \lam}  e^{\lam t} 
\ee
must be a solution to equations of motion at linearized level. Taking time derivative on both side of the above equation we find that $\ep$ satisfies the differential equation 
\be \label{kn2}
\p_t \ep = - f e^{\lam t} - {1 \ov \lam} \ka(\p_i) f e^{\lam t} = \tilde \lam (\p_i) \ep 
\ee
where 
\be 
\tilde \lam (\p_i) = \lam + \ka (\p_i)  \ .
\ee

In general different chaotic systems can have different $\ka(\p_i)$ and thus $\lam (\p_i)$.  
When transforming to Fourier space $\lam (k_i)$ can be interpreted as a momentum-dependent Lyapunov exponent, 
with the constant piece $\lam$ in $\tilde \lam (\p_i)$ as the Lyapunov exponent for variables without 
any spatial dependence.  
We will see later such a general shift symmetry could lead to a diffusive spreading around the exponential growth~\eqref{mm0} in coordinate space at large distances. 
As before due to that the action is invariant under the shift symmetry, the exponential mode is invisible to energy density and flux.

\section{Correlation functions in chaos EFT} \label{sec:coeft}

%In this section we study various near-equilibrium two-point functions of the hydrodynamical mode $\sig$ in the chaos EFT introduced in last section. They will be used in next sections for computing two-point functions of energy density and energy flux, and four-point functions of general operators. 
%In particular, we show that exponential growth and diffusion general leads to ballistic propagation and characterize the 
%class of theories for which~\eqref{btv} applies. 

In this section we study the near-equilibrium two-point  functions of the hydrodynamical mode $\sig$ in the chaos EFT. 
These functions will clarify the implications of the various shift symmetries that we introduced in the last section.  
They will also be used later for computing correlation functions of the energy density/energy flux, and four-point functions of general few-body operators.

We will denote the retarded, advanced and symmetric Green's functions of $\ep$ near equilibrium as 
$G_R, G_A, G_S$ respectively, which can be obtained from ($x= (t, x^i)$)
\be \label{hj1}
G_R  (x)  =i \vev{\ep (x) \ep_a (0)} \quad G_A  (x)  = i \vev{\ep_a  (x)  \ep (0)}  = G_R (-x)  , \quad G_S  (x)  =\vev{\ep(x)  \ep (0)} 
\ee
while the Wightman functions are\footnote{Recall that the subscript $1,2$ denotes the segment of the CTP an operator 
is inserted.}
\be \label{gpm}
G_+ (x) = \vev{\ep_2 (x) \ep_1 (0)} = G_S - {i \ov 2} G_R \; ({\rm for} \; t >0) , \quad
G_- (x) =  \vev{\ep_1 (x) \ep_2 (0)} = G_+ (-x) \ .
\ee
These functions can be obtained from path integrals of the quadratic action~\eqref{onbm}. In particular, the relations~\eqref{fdt1} 
ensure that they satisfy the fluctuation-dissipation relations  
\be\label{fdt2}
 G_S (x)=  - {i \ov 2}  \coth {i \b_0 \p_t \ov 2} (G_R  (x)- G_A (x)) \ .
 \ee
For a system with a shift symmetry, as discussed in last section, $\ep$ has an exponentially 
growing mode. The evaluation of $G_R, G_S$ from~\eqref{onbm} is subtle as one must be careful about
possible contributions at time infinities. 

To avoid too much technicality, we will proceed with a shortcut.  
The retarded Green's function $G_R (x) $ %= \vev{\ep (t) \ep_a (0)}_{\rm hydro}$ 
can  be obtained by inverting the different operator $K = \b_0 (\p_t f_1 + h_1 \p_i^2) \p_t$ in the $ \ep_a K \ep$ term in~\eqref{onbm}, with retarded boundary condition. More explicitly, 
\be \label{nem9}
 G_R (x) =- \int_C  {d^d k \ov (2 \pi)^d} {e^{- i \om t + i k_i x^i }\ov K }  \ , 
\ee
where to ensure the retarded boundary condition, the contour $C$ in the complex $\om$-plane should be chosen 
so that all the poles of integrand should lie below it. In particular, if $K$ has a zero in the upper half $\om$-plane, $C$ must be deformed to go above it. With $G_R$, one can then use~\eqref{fdt2} to obtain $G_S$, and then $G_\pm$ from~\eqref{gpm}. 
We will discuss the behavior of $G_R$ below leaving the expressions for $G_S$ and $G_\pm$ in Appendix~\ref{app:c}. 

We will assume that the system has a general shift symmetry~\eqref{kn} with~\eqref{kn1}, which includes point-wise shift~\eqref{ene} and the constant shift of a $(0+1)$-dimensional system as special cases. 

The presence of the shift symmetry~\eqref{kn} and equation~\eqref{kn2} implies that $f_1$ and $h_1$ in~\eqref{onbm} can be written in a form\footnote{Strictly speaking, invariance of~\eqref{onbm} only requires the combination $K$ to be proportional to $\p_t - \tilde \lam (\p_i)$.~\eqref{umm} follows by requiring the action in the presence of external fields also has the shift symmetry. See Appendix~\ref{app:a}.} 
\be \label{umm}
f_1 = (\p_t - \tilde \lam (\p_i^2)) a (\p_t, \p_i^2), \qquad h_1 = (\p_t - \tilde \lam (\p_i^2)) b (\p_t, \p_i^2) \ .
\ee
For later convenience we will denote 
\be \label{mio}
\sD  (\p_t , \p_i^2) \equiv - {b (\p_t, \p_i^2) \ov a (\p_t, \p_i^2)} = D_E+ O(\p_t, \p_i^2)
\ee
which can be interpreted as the ``diffusion'' operator, as its leading term in a derivative expansion gives the energy diffusion 
constant (see~\eqref{km}). Thus now the operator $K$ can be written in momentum space as 
\be \label{nem1}
K =i  \b_0 \om a(\om, k) (\om - i \tilde \lam (k)) \le(\om + i \sD (\om, k) k^2 \ri) , \quad k^2 = k_i^2  \ 
\ee
and~\eqref{nem9} becomes 
\be \label{nem}
 G_R (x) ={i \ov \b_0} \int_C  {d^d k \ov (2 \pi)^d} {e^{- i \om t + i k_i x^i }\ov \om a(\om, k) (\om - i \tilde \lam (k)) \le(\om + i \sD (\om, k) k^2 \ri)  }  \ .
\ee
Note that for point-wise shift symmetry we simply have $\tilde \lam = \lam$ and the expression for $(0+1)$-dimension
is obtained by setting all $k_i$ to zero.  

For simplicity we will assume that for {\it real} $k$, $K$ does not have any 
zero on the upper $\om$-plane other than $\om = i \tilde \lam$. Other zeros of $K$ on upper half $\om$-plane 
may indicate instabilities or secondary Lyapunov exponents if they arise due to some other shift symmetries.
 In the lower half plane we expect there to be additional zeroes (for real $k$) in $K$ corresponding to solutions of\footnote{In principle 
 $a(\om, k)$ may also have zeros.} 
\be \label{dispersion}
\om + i \sD (\om, k) k^2  = 0,
\ee
As we will show explicitly in Sec \ref{sec:pole} solutions to \eqref{dispersion} give rise to poles in the two-point functions of energy density and energy flux. In particular, in the limit of small $\omega, k$ they include the standard energy diffusion pole 
\be
\label{diff}
\omega = - i D_E k^2 + \dots
\ee
and (often infinite) other quasinormal poles 
with schematic dispersion relations at small $k$ 
\be 
\om = \om_0 + O(k^2) , \quad  \om_0 \sim O(\b_0), \quad {\rm Im} \, \om_0 < 0 \  .
\ee
%Note that outside the small frequency limit such dispersion relations can include potentially large corrections to pure diffusion. 

Now let us examine the behavior of~\eqref{nem} for various situations: 

\ben 

\item $(0+1)$-dimensional chaotic systems. In this case we have $K = i \b_0 \om^2 a(\om) (\om - i \lam)$ which gives 
\be \label{hnk00}
G_R (t) = \th (t) \le(c e^{\lam t} + \cdots
 %\tilde f_R (t)
  \ri), \qquad c = - \frac{1}{\b_0 a (i \lam) \lam^2} 
\ee
where the exponential term comes from evaluating the pole at $\om = i \lam$ and $\cdots$ denotes the remaining (non-exponentially growing) contributions. 

\item Point-wise shift symmetry.  In this case we again have a pole at $\om = i \lam$, whose contribution gives 
\be \label{jk1}
G_R = - \th (t) {e^{\lam t} \ov \b_0 \lam}  \int {d^{d-1} k \ov (2 \pi)^{d-1}} \,  {e^{ i k_i x^i} \ov  (\lam  + k^2 \sD (i \lam, k^2)) a (i\lam, k^2)} \ .
\ee
In general $\sD (i \lam, k^2)$ is a complicated function of $k^2$. Let us first consider a special case in which 
\be \label{xd}
\sD (\om, k^2) = D_E = {\rm const}
\ee
with $D_E$ the energy diffusion constant, which happens for example for the SYK chain discussed in Sec.~\ref{sec:sykc}.  

In this case we have $h_1 = D_E f_1$ and $G_i = - D_E \p_i H$ and hence a diffusion pole satisfying \eqref{diff} to all orders in derivatives. Now the integrand of~\eqref{jk1} has a pole 
at  $k^2 =  - {\lam \ov D_E}$, whose contribution gives 
\be \label{balk}
G_R = c \th (t) e^{\lam (t - {|\vec x| \ov v_B} )} + \cdots, \qquad v_B^2 = \lam D_E %\qquad C = {1 \ov K_H (i \lam, - {\lam \ov D_0})} \ .
\ee
with $c$ some constant. We therefore see that ballistic propagation of chaos can arise from the combination of an exponentially growing mode and the spatial propagation of diffusion. 
 
For a general $\sD (\om, k^2)$ we can get the same behavior if we suppose that the equation
\be \label{nnp}
\lam + k^2 \sD (i \lam, k^2)  = 0
\ee
has a solution at some $-k_C^2 < 0$. Then we get ballistic behaviour with a butterfly velocity given by
\be \label{by}
v_B^2 = \frac{\lambda^2}{k_C^2} 
\ee
Note that the fact $v_B$ is coming from a solution to \eqref{dispersion} means it is again determined by a pole in the energy density two point function. Specifically \eqref{nnp} implies that to get ballistic behaviour this correlation function should have a pole which eventually crosses the point $\omega(k) = i \lambda$ for some imaginary wave-vector $k = i k_C$. The value of the wavevector $k_C$ at which this happens determines $v_B$ according to 
\begin{equation}
\label{relation2}
\omega(i k_C) = i \lambda \;\;\;\;\;\;\;\;\;\;  v_B = \frac{\lambda}{k_C} 
\end{equation}
In the case where the pole that determines $v_B$ is exactly diffusive this gives the simple relation $D_E = {v_B^2 \ov \lam}$ seen above and in examples of SYK chains. More generally, the pole that determines $v_B$ will obey some more complicated dispersion relation given by solving \eqref{dispersion}. Nevertheless, our approach predicts that the relationship \eqref{relation2} between this pole and the butterfly velocity will hold.

\item General $\tilde \lam (k)$.  Performing the $\om$ integral in~\eqref{nem} we find that 
\be \label{hjj}
G_R = - \th (t) {1 \ov \b_0} \int {d^{d-1} k \ov (2 \pi)^{d-1}} \,  {e^{\tilde \lam (k) t + i k_i x^i}  \ov  \tilde \lam (\tilde \lam  + k^2 \sD (i \tilde \lam, k^2)) a (i \tilde \lam, k^2)}  \ .
\ee 
We again have the behavior~\eqref{balk} if\footnote{For the discussion below to hold we need $\tilde \lam (k)$ to behave sufficiently well at large $k$ so that contour integrations can be performed.} 
\be \label{kkp}
\tilde \lam  + k^2 \sD (i \tilde \lam, k^2) = 0
\ee 
has a solution at some $- k_C^2 < 0$. Now define 
\be 
\bar \lam = \tilde \lam (- k_C^2) , \qquad v_B =  {\bar \lam \ov k_C} 
\ee 
then we have 
\be \label{balk1}
G_R = c \th (t) e^{\bar \lam (t - {|\vec x| \ov v_B} )} + \cdots \ .
\ee
Note that $\bar \lam$ is in general different from $\lam = \tilde \lam (k=0)$ and may be interpreted as an ``averaged'' Lyapunov exponent over different $k$. 

Equation~\eqref{hjj} can also have a different regime. 
Suppose for small $k$, 
\be 
\tilde \lam (k) = \lam - D_0 k^2 + \cdots 
\ee
with some  $D_0 > 0$. Then for large $|\vec x|$ and $\sqrt{\lam D_0} t \gg |\vx|$, the integrals are dominated by small $k$ region, and 
we can evaluate~\eqref{hjj} by saddle point of the exponent, which gives
\be \label{diffchaos}
G_R \sim \th (t) {1 \ov t^{d-1 \ov 2}} e^{\lam t - {\vec x^2 \ov 4 D_0 t}} + \cdots \ .
\ee 
Note that this regime is qualitatively different to the ballistic propagation in~\eqref{balk} or~\eqref{balk1} which arose from the interplay of the exponential growing mode and other poles (such as energy diffusion). In contrast the entire functional form of \eqref{diffchaos} is determined by the shift symmetry $\lambda(k)$ alone. 
 
\een

\section{Phenomenon of  pole skipping} \label{sec:pole}

In this section we examine two-point functions of the energy density $\sE$ and flux $\sJ^i$ to all derivative orders. Their behavior confirms our earlier expectation that the exponentially growing mode should be invisible to them. 
Nevertheless, we will find that the hydrodynamic origin of chaos predicts the phenomenon of  ``pole-skipping'' in the response functions of the energy density and flux, as indicated in Fig.~\ref{fig:vanishingpole}. Remarkably this phenomenon implies that both the Lyapunov exponent and the butterfly velocity can be extracted from the energy density two-point function alone.

We will denote the retarded, advanced and symmetric Green's functions for $\sE, \sJ^i$ as $\sG_R, \sG_A, \sG_S$ respectively,  For example, various density-density functions are obtained from 
\be  \label{hj2}
\sG_R^{\sE \sE} (x) = i \vev{\sE_r (x) \sE_a (0)} , \quad 
\sG_A^{\sE \sE} (x) = i \vev{\sE_a (x) \sE_r (0)} , \quad
\sG_S^{\sE \sE} (x) = i \vev{\sE_r (x) \sE_r (0)} 
\ee
where the explicit forms of the symmetric $\sE_r$ and antisymmetric part $\sE_a$ of the energy density 
are given in Appendix~\ref{app:a}. One finds, for example, various retarded functions are given by 
\bega \label{u1}
\sG_R^{\sE \sE} (x) =  \b_0^2  f_1 h_1 \p_t \p_i^2 G_R (x)  , \qquad
\sG_R^{\sE \sJ^i} =  -\b_0^2  f_1 h_1 \p_t^2 \p_i G_R , \\ \sG_R^{\sJ^i\sE} =  - \b_0^2   h_1 f_1  \p_t^2  \p_i G_R  , \qquad
\sG_R^{\sJ^i\sJ^j} =  - \b_0^2   h_1^2 \p_t^2 \p_j  \p_i G_R  \ 
\end{gather} 
where $G_R$ is retarded function of $\sig$  given in~\eqref{nem}.  In the above expressions we have suppressed various ``contact'' terms which are given explicitly in Appendix~\ref{app:a}. One can also check that with $G_R, G_S$ satisfying~\eqref{fdt2}, various $\sG_R$ and $\sG_S$ satisfy the fluctuation-dissipation relations
\be 
\sG_R (x) - \sG_A (x) = 2 i \tanh{i \b_0 \p_t \ov 2} \sG_S (x) \ .
\ee

We now use $\sG_R^{\sE \sE} (x)$ as an illustration for the pole-skipping phenomenon with parallel discussions for others. From~\eqref{nem}-\eqref{umm} and~\eqref{nem1} in~\eqref{u1} we find in momentum space  
\be \label{eer}
\sG_R^{\sE \sE} (\om, k) =  \b_0  {(\om - i \tilde \lam (k))k^2 b(\om, k) \ov \om + i \sD (\om, k) k^2}  \ . 
\ee
Notice that the factor $\om - i \tilde \lam (\om,k)$ now appears in the upstairs. Thus there is no exponential behavior. 
Let us first look at the~\eqref{eer} to leading order in the small $\om, k$ limit. Comparing with~\eqref{ing} we identify ${\ka \ov \b_0^2} = \tilde \lam (k=0) b (\om=k=0)$, and using~\eqref{km},~\eqref{mio} we find~\eqref{eer} becomes  
\be \label{smk}
\sG_R^{\sE \sE} (\om, k)= - {c_0 \ov \b_0} {D_E k^2 \ov - i \om + D_E k^2}  
\ee
which is the standard form.

Now due to the presence of  the factor $\om - i \tilde \lam (\om,k)$ in the upstairs of~\eqref{eer} there is a new phenomenon. 
Equation~\eqref{eer} has a pole at 
\be 
\om = - i \sD (\om, k^2) k^2 
\ee
which for small $\om, k$ is simply the standard diffusion pole $\om = - i D_E k^2$ as exhibited in~\eqref{smk}. Consider continuously changing the value of $k$ 
until $k = \pm i k_C = \pm i {\bar \lam \ov v_B} $ which satisfies~\eqref{kkp}. At this value of $k$, $\om = i \bar \lam$ 
and thus the zeros in upstairs and downstairs of~\eqref{eer} precisely coincide and cancel each other. 
We then find a line of poles which suddenly skips at that point, as indicated in Fig.~\ref{fig:vanishingpole}. 

Note that this phenomenon is a general consequence of formulating a hydrodynamic theory of chaos with a shift symmetry $\tilde \lam (k)$. For the case of a point-wise shift symmetry, we expect that the pole-skipping will always occur at a frequency $\bar \lam = \lam$, while for the extreme diffusion case we further have that the $k_C  = \sqrt{\lam \ov D_E}$. However note that in all cases we can use the location of this pole-skipping to read off both the Lyapunov exponent and butterfly velocity from a computation of $\sG_R^{\sE \sE} (\om, k)$ alone. 

This phenomenon has implicitly been present in several previous calculations of the energy two-point function in chaotic theories. For instance it can be seen in the expression found for a SYK chain (equation (4.15) of~\cite{Gu}) whose analytic continuation to Lorentzian signature gives (using our notations)
\be 
\sG_R^{\sE \sE} (\om, k) = {\sN c_0 \ov \b_0^2} {i \om \le( {\om^2 \ov \lam_{\rm max}^2} + 1\ri) \ov - i \om + D_E k^2}  \ .
\ee 
As we discussed earlier the SYK chain of~\cite{Gu} has point-wise shift symmetry and is an example of extreme diffusion (see Sec.~\ref{sec:sykc}). The same pole-skipping phenomenon has also been observed in a momentum conserved system in~\cite{Saso} at precisely the value stated above. This strongly suggests that not only is this phenomenon expected to hold for systems with full energy-momentum conservation, but also the locations of pole-skipping remain the same.

\section{Correlation functions of general few-body operators}\label{sec:corre}

In Sec.~\ref{sec:cEFT} we proposed a chaos EFT in which the hydrodynamic variable corresponding to energy conservation has an exponentially  growing mode. We showed in Sec.~\ref{sec:coeft}  that this mode has ballistic propagation as in~\eqref{iem1}
or diffusive spreading around the exponential growth~\eqref{mm0}. We also saw that such exponential behavior does not show up in energy density or energy flux, or their corresponding correlation functions, but does give rise to the phenomenon of pole-skipping.  In this section we discuss their relevance for correlation functions of general few-body operators. 

There are two key aspects we would like to elucidate. Firstly, despite that correlation functions 
such as~\eqref{iem} require a contour with at least four segments, to leading order in the large $\sN$ limit 
such four-point functions are in fact controlled by two-point near-equilibrium functions of the hydrodynamical  variable $\sig$
discussed in Sec.~\ref{sec:coeft}. Secondly, with the imposing of a shift symmetry in the couplings between a general few-body operator 
and $\sig$, the exponentially growing mode does not affect TOCs~\eqref{TOC}, but does 
show up in OTOCs~\eqref{OTOC}, resulting in~\eqref{iem0}--\eqref{iem1}.

%why should a solution to be the classical equations of motion be relevant for calculating out-of-time ordered correlation functions, which need to be calculated using a four-fold Schwinger-Keldysh contour.  Secondly, we need to explain why, if such a mode exists on a single SK contour, it does not also appear in time-ordered correlation functions. 

\subsection{General structure} 

Let us first discuss the general structure of the couplings of a general few-body operator to $\sig$. 
 To simplify our discussion we will focus on studying OTOCs in $0+1$ dimensional systems. We then briefly discuss how the basic structure could be generalised to higher dimensional systems with strong momentum dissipation.

We imagine each few-body operator $V(t)$ can be separated into a bare operator $\hat V (t) $ dressed by a ``hydrodynamical'' 
cloud as indicated in Fig.~\ref{fig:hydrocloud}. The bare operators can only communicate with themselves, i.e. 
$\vev{\hat V \hat W} =0$ for generic $V \neq W$. %and thus nontrivial correlations between $V$ and $W$ are through their 
%hydrodynamical clouds. 
Such a separation clearly makes sense only in the limit of large number of degrees of freedom. 
% e.g. 
%very natural in the gravity description of a holographic theory.\footnote{\textcolor{blue}{I am a bit nervous about saying this. Clearly in holographic theories other interactions are possible beyond hydro. We just are not describing them here.}}
More explicitly, we can expand $V (t)$ in power series of $\ep (t) = \sig (t) - t $ as 
\be \label{cop}
V (t) = \hat V (t) + L^{(1)} [\hat V \ep ] (t)  + O(\ep^2)   %\le(1 + f^{(1)} \ep + O(\ep^2) + \cdots \ri)
\ee
where  
%we have used a schematic notation. 
$L^{(1)}$ is a differential operator acting on both $\hat V$ and $\ep$, %should More explicitly, the term $\hat V f^{(1)} \ep $ 
and  should be understood as 
\be \label{uen}
L^{(1)} [\hat V \ep ] = \sum_{n,m=0}^\infty c_{nm} \p_t^n \hat V \p^m_t \ep
\ee
where $c_{mn}$ are constants. In other words we take the most general possible local coupling between $\hat V$ and $\ep$.
One expects that for a chaotic system $L^{(1)}$ should not depend on the specific form of $ V$, and only on some gross features such as spin or scaling dimension.
The couplings which are quadratic and higher in $\ep$ will be neglected as due to~\eqref{scad} they will give subleading corrections in $1/\sN$.%\textcolor{blue}{Equivalently we are looking at physics before the scrambling time?}

 An example of~\eqref{uen} is SYK or holographic AdS$_2$ theories where 
the full dressed operator $V$ has the form~\cite{MSY, Jensen, MS}
\be \label{uen1}
 V (t) = (\p_t \sig)^{\De_V} \hat V (\sig (t))
\ee
where $\De_V$ is a constant given by the infra-red scaling dimension of the operator ${V}$. Equation~\eqref{uen1} has a simple geometric interpretation: the bare operator $\hat V$ 
is simply the pull-back of $V$ to the fluid spacetime. Expanding~\eqref{uen1} in $\ep$ we find that 
\be \label{pom}
 V(t) = \hat V (t) + \le(\De_V \ep' (t) + \ep (t) \p_t \ri) \hat V (t) + \cdots  \ .
\ee

Now consider a general $4$-point function between two few-body operators $\hat V$ and $\hat W$ ordered in a certain way 
\ie \label{npoi}
G_{i_1 i_2 i_3 i_4} (t_1, t_2, t_3, t_4) =  {\vev{\sP V_{i_1} (t_1) V_{i_2}  (t_2)  W_{i_3} (t_3) W_{i_4} (t_4) }  \ov 
\vev{\sP V_{i_1} (t_1) V_{i_2} (t_2)} \vev{\sP W_{i_3} (t_3) W_{i_4} (t_4)}} 
\fe
where $\sP$ denotes path ordering along the contour in the right plot of Fig.~\ref{fig:CTPcontour} and $i_{1,2,3,4} $ (taking values from $1$ to $4$)  denote on which contour each operator is inserted. Note that $V_{i1} (t) =\hat V_{i1} (t) +  L^{(1)} (\hat V_{i1} \ep_{i1}) + O(\ep_{i1}^2)$
where $\ep_{i1} = \sig_{i1} (t) - t$ and $\sig_{i1} (t)$ is the inverse of $X_{i1} (\sig)$ discussed in Sec.~\ref{sec:eng}. Now since there are four segments we need to introduce four $X$'s or correspondingly four $\sig$'s depending on whether one wants to write down the action in the fluid or physical spacetime.

Inserting~\eqref{cop} into~\eqref{npoi} and keeping keeping in mind there is no correlation between $ \hat V$ and $\hat W$, we find that to leading order in
large $\sN$ the four-point function reduces to various two-point functions of $\ep$, 
\be \label{onne}
G_{i_1 i_2 i_3 i_4} - 1=  \vev{B_V^{i_1 i_2} (t_1, t_2) B_W^{i_3 i_4} (t_3, t_4)}
\ee
with 
\be \label{njm}
B_V^{i_1 i_2} = {1 \ov g_V}  \le(L^{(1)}_{t_1}  [g_V (t_{12})  \ep_{i_1} (t_1)]  + L^{(1)}_{t_2} [g_V (t_{12}) \ep_{i_2} (t_2)]  \ri)
\ee
where the subscript on $L^{(1)}$ denotes that which variable it acts on, and 
\be \label{gvt}
g_V (t_{12}) \equiv \vev{\sP \hat V_{i_1} (t_1) \hat V_{i_2} (t_2)}, \qquad t_{12} = t_1 - t_2 \ .
\ee
$B_W$ is similarly defined with $V$ replaced by $W$. $B_V$ can be considered as defining the effective vertex 
for $\hat V \hat V$ coupling to $\ep$. Since two-point functions on a four-segment contour reduce to those on a CTP 
contour (the left plot of Fig.~\ref{fig:CTPcontour}), we conclude that  regardless of the orderings,~\eqref{onne} can be computed using correlation functions of $\sig$ on a CTP contour discussed in Sec.~\ref{sec:coeft} and Appendix~\ref{app:c},
 as indicated in Fig.~\ref{fig:otoreduction}.  This discussion also makes clear that at this order in $1/\sN$ 
the only interesting correlation functions between $V$ and $W$ are four-point functions as all 
higher-point functions reduce to them.

\begin{figure}
\begin{center}
\resizebox{80mm}{!}{\includegraphics{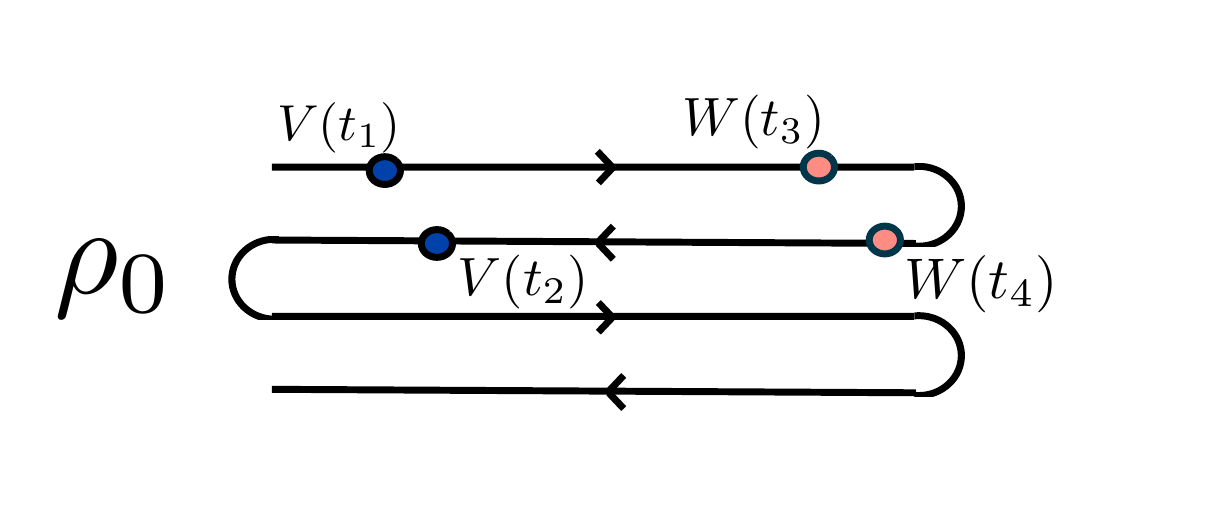}}
\resizebox{80mm}{!}{\includegraphics{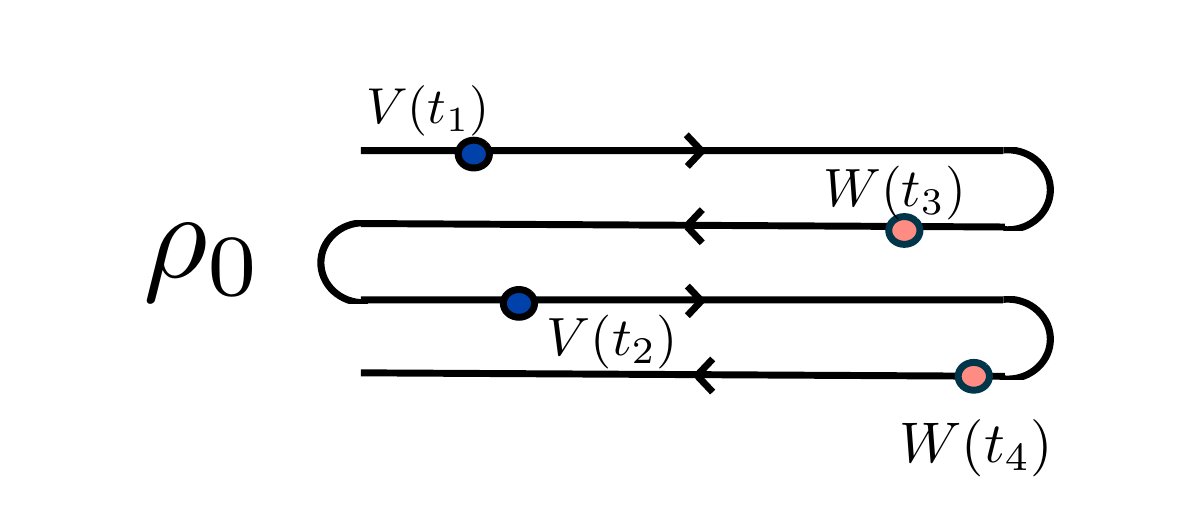}}
\caption{Left: operator insertions for~\eqref{o1}. Right: operator inserations for~\eqref{o2}.
}
\label{fig:4contours}
\end{center}
\end{figure}

%

%To summarize, from~\eqref{onne} we find that no matter what orderings 
%of $V$ and $W$'s are, to leading order their correlation functions are always controlled by various two point functions of the hydrodynamic fields. 

Let us consider two explicit examples indicated in Fig.~\ref{fig:4contours}. The first is TOC as in~\eqref{TOC}
\be\label{o1}
G_4(t_1,t_2,t_3,t_4) = \langle \sP  V_1(t_1) V_2(t_2) W_1(t_3) W_2(t_4) \rangle  %\ov \vev{V (t_2) V (t_1)} 
= \vev{V(t_2) W(t_4) W(t_3) V(t_1)}  
\ee
and the second is the OTOC~\eqref{OTOC} 
\be \label{o2}
H_4(t_1,t_2,t_3,t_4) = \langle \sP V_1(t_1) V_2(t_2) W_1(t_3) W_3 (t_4) \rangle  
= \vev{W (t_4) V (t_2) W (t_3) V (t_1)} 
\ee
with $t_{3,4} \gg t_{1,2}$, where in the second equalities of~\eqref{o1}--\eqref{o2} we have given the specific 
orderings of various operators. In these definitions for notational simplicity we have suppressed the downstairs of~\eqref{npoi}. 
Using~\eqref{onne} we find the difference between the two 
\be \label{mm1}
H_4 - G_4 = {1 \ov g_V g_W}  L^{(1)}_{t_2}  L^{(1)}_{t_4} [g_V(t_{12}) g_W (t_{34}) \De (t_{42}) ], \qquad \De (t_{42} ) = \vev{[\ep (t_4) , \ep (t_2)]} \ .
\ee
Thus if the couplings $c_{mn}$ and two-point functions of $\hat V, \hat W$ are such that the exponential mode does not appear in ordered four-point function $G_4$, they will always appear in out-of-time order four-point functions $H_4$.

\subsection{Shift symmetry for effective vertex} 

We will now require the effective vertex~\eqref{njm} respect the shift symmetry~\eqref{uie}, i.e. it should be invariant 
under $\ep_i \to \ep_i + c e^{\lam t}$ with $c$ some constant. This implies that 
\be \label{om}
 L^{(1)}_{t_1} [g_V  (t_{12})  e^{\lam t_1}]   + L^{(1)}_{t_2} [g_V  (t_{12})  e^{\lam t_2}] = 0
 \ee
which implies that % \textcolor{blue}{(note change in $F$ notation to show how condition depends on $\lambda$)}
\be \label{mme}
{F_{\rm even} (\lambda, t) \ov F_{\rm odd} (\lambda, t)} = - \tanh {\lam t \ov 2} \ .
\ee
In the above equations we have introduced 
\be 
F_{\rm even} (\lambda, t) = \sum_{n \; {\rm even}} f_n (\lam) \p^{n}_t g_V (t) ,\qquad 
F_{\rm odd} (\lambda, t) = \sum_{n \; {\rm odd}} f_n (\lam) \p^{n}_t g_V (t)  \ .
\ee
where 
\be 
f_n = \sum_m c_{nm} \lam^m  \ .
\ee
Note that with equation~\eqref{mme} we can write $L^{(1)}_{t_1} [g_V  (t_{12})  e^{\lam t_1}]  =- L^{(1)}_{t_2} [g_V  (t_{12})  e^{\lam t_2}] $ as 
\be \label{jme}
L^{(1)}_{t_1} [g_V  (t_{12})  e^{\lam t_1}] = e^{\lam t_1}  \le(F_{\rm even} (\lambda, t_{12})  + F_{\rm odd} (\lambda, t_{12})\ri)
=- {F_{\rm even} (\lam, t_{12}) \ov \sinh {\lam t_{12} \ov 2}} e^{\ha \lam (t_1 + t_2)}  \ .
\ee 
For $V$ a Hermitian operator, we have $g_V^* (t) = g_V (-t)$ and thus $F_{\rm even}^* (\lam, t)= F_{\rm even} (\lam, -t)$ and 
$F_{\rm odd}^* (\lam, t)= - F_{\rm odd} (\lam, -t)$.

While is does not naturally follow from our logic, we will see in next subsection that in order for the exponential growing behavior to cancel for all the TOCs, we also need to impose the version of~\eqref{om} with $\lam \to -\lam$, i.e. 
\be \label{omm2}
 L^{(1)}_{t_1} [g_V  (t_{12})  e^{-\lam t_1}]   + L^{(1)}_{t_2} [g_V  (t_{12})  e^{-\lam t_2}] = 0
 \ee
which implies 
\be \label{mme1}
{F_{\rm even} (-\lambda, t) \ov F_{\rm odd} (-\lambda, t)} =  \tanh {\lam t \ov 2} \
\ee
and 
\be \label{jme1}
L^{(1)}_{t_1} [g_V  (t_{12})  e^{-\lam t_1}] %= e^{\lam t_1}  \le(F_{\rm even} (\lambda, t_{12})  + F_{\rm odd} (\lambda, t_{12})\ri)
=  {F_{\rm even} (-\lam, t_{12}) \ov \sinh {\lam t_{12} \ov 2}} e^{-\ha \lam (t_1 + t_2)}  \ .
\ee 

Now let us consider applying~\eqref{mme} to the specific example~\eqref{pom} for SYK and AdS$_2$. We find that  
\be
\label{sykcond}
{\p_t g_V \ov g_V} = - {\De_V \lam_0 \ov \tanh {\lam_0 t \ov 2}}, \qquad \lam_0 = {2 \pi \ov \b_0} 
\ee
which can be integrated to give
\be \label{gvb}
g_V = {c \ov \le(\sinh  {\lam_0 t \ov 2} \ri)^{2\De_V}} \ .
\ee
The above expression is precisely the expected behavior in SYK and AdS$_2$ for two-point function of $\hat V$. 
Note that~\eqref{gvb} also satisfies~\eqref{omm2}.

\subsection{Ordered and out-of-time ordered four-point functions} 

We will now show that the shift symmetry~\eqref{om} implies that the exponential behavior is canceled for the
ordered four-point function~\eqref{o1}, but leads to chaotic behavior~\eqref{iem0} for OTOC~\eqref{o2}. 

More explicitly, from~\eqref{onne}--\eqref{njm} we find~\eqref{o1} can be written as
\be  \label{timeordered}
G_4 - 1= {1 \ov g_V g_W} \le( L^{(1)}_{t_1}  L^{(1)}_{t_3}  G_+ (t_{31})  + L^{(1)}_{t_1}  L^{(1)}_{t_4} G_+ (t_{41}) +
L^{(1)}_{t_2}  L^{(1)}_{t_3} G_- (t_{32}) +  L^{(1)}_{t_2}  L^{(1)}_{t_4} G_- (t_{42})  \ri) g_V g_W 
\ee
where in the above equation it should be understood that $L^{(1)}$'s also act on $g_V (t_{12})$ or $g_W (t_{34})$  after them. 
The contribution from the exponential modes to the above equation can be extracted using the hydrodynamic Green's functions for $t > 0$. First let us consider a non-maximally chaotic theory, for which we have (see Appendix~\ref{app:c})
\be
G_+(t) = c_+ e^{\lambda t} + \cdots \;\;\;\;\;\;\; G_{-}(t) = c_- e^{\lambda t} + \cdots %\;\;\;\;\; c = i (c_+ - c_-) 
\ee
with $c_\pm$ some constants. 
With some manipulations it can be shown that the contribution of the exponential mode to~\eqref{timeordered} can be written as
\be \label{nm}
G_4 - 1= B_W [\tilde \ep (t_3) , \tilde \ep (t_4) ] + \cdots 
\ee
where
\be 
\tilde \ep (t_3) = c_{12} e^{\lam t_3}, \quad \tilde \ep (t_4) = c_{12} e^{\lam t_4}, \quad
c_{12} = c_+ L^{(1)}_{t_1} [g_V e^{- \lam t_1}] +  c_- L^{(1)}_{t_2} [g_V e^{- \lam t_2}] \ .
\ee
Equation~\eqref{om} then implies that the contribution from the exponential mode precisely cancels out of this time ordered configuration. Note the above expression~\eqref{nm} has a natural interpretation in terms of a saddle point analysis. Namely, we can view the initial $V$ operators as sourcing a classical solution with an exponential mode $\epsilon_1 = \epsilon_2 = c_{12} e^{ \lam t}$ turned on in between the two operator insertions. The four point function $G_4$ then reduces to evaluating the effective vertex $B_W$ in this configuration, which vanishes due to condition~\eqref{om}.

\begin{figure}
\begin{center}
\resizebox{150mm}{!}{\includegraphics{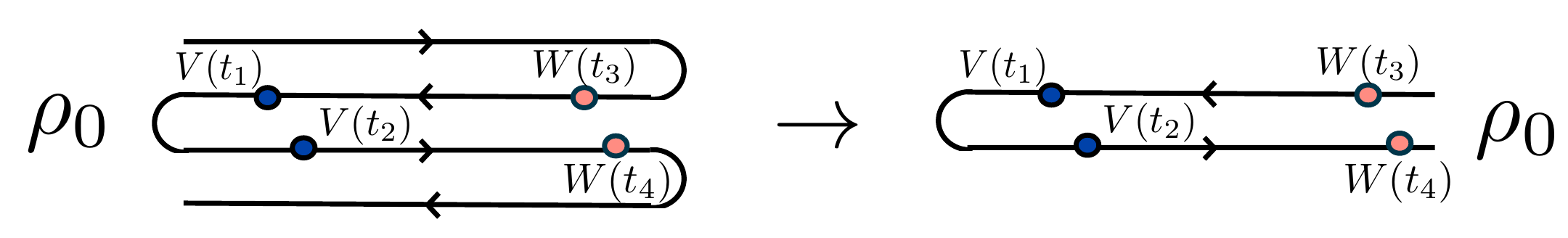}}
\caption{The configuration $\tilde{G}_4$ can be reduced to a four point function on a time reversed CTP contour.}
\label{fig:reversecontour}
\end{center}
\end{figure}

 {Note that whilst the condition \eqref{om} is sufficient to get cancellation of the exponential mode in \eqref{timeordered}, it is not enough to guarantee that the exponential growth cancels for all relevant time-ordered configurations. This can be seen by repeating the above analysis for the correlation function} 
\be
\label{to2}
\tilde{G}_4(t_1,t_2,t_3,t_4) = \langle \sP  V_2(t_1) V_3(t_2) W_2(t_3) W_3(t_4) \rangle  %\ov \vev{V (t_2) V (t_1)} 
= \vev{W(t_4) V(t_2) V(t_1) W(t_3)}   \ .
\ee
Notice that in contrast to~\eqref{o1}, this correlation function corresponds to operator insertions on a closed time path with $\rho_0$ imposed at $t = +\infty$. See Fig.~\ref{fig:reversecontour}.
%
%
%\textcolor{blue}{which can be written as }
%
%\be  \label{timeordered2}
%\tilde{G}_4 - 1= {1 \ov g_V g_W} \le( L^{(1)}_{t_1}  L^{(1)}_{t_3}  G_- (t_{31})  + L^{(1)}_{t_1}  L^{(1)}_{t_4} G_+ (t_{41}) +
%L^{(1)}_{t_2}  L^{(1)}_{t_3} G_- (t_{32}) +  L^{(1)}_{t_2}  L^{(1)}_{t_4} G_+(t_{42})  \ri) g_V g_W 
%\ee
%
%
{For the exponential modes to vanish in \eqref{to2} we find we now need to impose} the condition~\eqref{omm2}
%
%\be \label{om2}
% L^{(1)}_{t_1} [g_V  (t_{12})  e^{-\lam t_1}]   + L^{(1)}_{t_2} [g_V  (t_{12})  e^{-\lam t_2}] = 0
% \ee
%
%
{which corresponds to requiring invariance of the effective vertex~\eqref{njm} under an exponentially decaying mode $\ep_i \to \ep_i + c e^{-\lam t}$. Such a condition is natural for the correlation function~\eqref{to2} since this is a time reversed configuration relative to \eqref{timeordered}.} This gives an extra constraint~\eqref{mme1} on the couplings.
% corresponding to~\eqref{mme} with $\lambda \to -\lambda$. This condition can be seen to be satisfied by the couplings \eqref{sykcond} in the SYK model.

%
%

For the maximally chaotic chaos, a slightly more sophisticated analysis is required since (as discussed in Appendix~\ref{app:c}) the hydrodynamic Green's functions have additional terms. Namely, we have
\be
G_+(t) = ( c t + c_+) e^{\lambda t},  \;\;\;\;\;\;\; G_{-}(t) = (c t + c_-) e^{\lambda t}, \;\;\;\;\; c = - i  \b_0 (c_+ - c_-)  \ .
\ee
The terms proportional to $c_+,c_-$ will again vanish out of~\eqref{timeordered} if we impose the condition~\eqref{om}. To check that the $t e^{t}$ terms vanish we can write these terms as $G_+ = G_- = \partial ( c e^{\lambda_1 t}) / \partial \lambda_1$ and then set $\lambda_1 = \lambda$. Then we can then write the contribution of these terms to~\eqref{timeordered} as 
\be \label{tomax}
G_4 - 1= \frac{ \partial B_W [\tilde \ep (t_3; \lam_1) , \tilde \ep (t_4; \lam_1) ]  }{\partial \lambda_1}\bigg|_{\lambda_1 = \lam}
\ee
where
\be 
\tilde \ep (t_3; \lam_1) = c_{12}(\lambda_1) e^{\lambda_1 t_3}, \quad  \tilde \ep (t_4; \lam_1) = c_{12}(\lambda_1) e^{\lambda_1 t_4} , \quad c_{12}(\lambda_1) =  c L^{(1)}_{t_1} [g_V e^{- \lambda_1 t_1}] +   c L^{(1)}_{t_2} [g_V e^{- \lambda_1 t_2}] \ .
\ee
After imposing~\eqref{om} we find that there will be an exponentially growing contribution in~\eqref{tomax} unless $c_{12}(\lambda) = 0$. However this is precisely satisfied if we use the fact that the effective vertex should also be invariant under \eqref {omm2}.\footnote{We thank Ping Gao for pointing this out to us.} The fact that we need both conditions to cancel the exponential growth in~\eqref{timeordered} alone is 
interesting.%surprising, and its physical origin is not currently clear to us}  (\HL{What happens to~\eqref{to2} for maximal chaos?})

Since the shift symmetry of the coupling ensures the exponential modes do not appear in the TOC then they will appear instead from~\eqref{mm1} in the OTOC. We then find the contribution of the exponential mode to the OTOC is given by 
\be 
H_4 -1 =   - { i c  \ov g_V g_W}  L^{(1)}_{t_2}  L^{(1)}_{t_4} [g_V(t_{12}) g_W (t_{34}) e^{\lam t_{42} }] + \cdots \ .
\ee
Using equations~\eqref{jme} and~\eqref{jme1} it is possible to write the above equation in a form which separates the dependence on 
the relative time $t_{12}$ ($t_{34}$) and ``center of mass time'' $\ha(t_1 + t_2)$ ($\ha(t_3 + t_4)$) for $V$'s ($W$'s)
\begin{equation}
\label{OTOCexp}
H_4 -1 = \frac{i c}{g_V g_W}  \frac{F_{\mathrm{even}}(-\lambda, t_{12} )} {\mathrm{sinh}{\lambda t_{12} \ov 2}}\frac{{\tilde F}_{\mathrm{even}}(\lambda, t_{34} )}{\mathrm{sinh}{\lambda t_{34} \ov 2}} e^{\lambda {(t_3 + t_4- t_1 - t_2 )}/2} + \cdots \ .
\end{equation}
where $F$ refers to the coupling of $\epsilon$ to $V$ and $\tilde{F}$ refers to the coupling of $\epsilon$ to $W$. Note for maximal chaos this is consistent with the functional form for OTOCs proposed in \cite{Kitaev}.

In the case of the SYK model in the infra-red limit we know the exact couplings and~\eqref{mme} is satisfied as a result of the $SL(2,R)$ symmetry of generic correlation functions. To compare with the results of~\cite{MSY},  let us consider the full $G_4$ 
(not just the exponential parts) for the example of~\eqref{nd3} at maximal chaos. Using the explicit expression~\eqref{gpp} for $G_+$ obtained in Appendix~\ref{app:c}, we find that (in the equations below we have set $a_2 = 1$, $\b_0 = 2 \pi$ and  $\lam_{\rm max} =1$)
\be 
G_4 -1 = {\Delta_V \Delta_W \ov 2 \pi } \le[\le(- 2 + { t_{12} - 2 \pi i \ov  \tanh {t_{12} \ov 2}}
     \ri) \le(-2 + {t_{34} \ov \tanh {t_{34} \ov 2}}\ri) \ri]
\ee
which agrees fully with Lorenzian continuation of (4.30) of~\cite{MSY}.\footnote{The additional $ -2 \pi i$ term compared with (4.30) of~\cite{MSY} is due to the fact we are considering a slightly different ordering. Here we consider $VWWV$ while there $VVWW$ is computed: the two orderings can be related by taking $t_1 \to t_1 + i \b_0$.}  Similarly, for OTOC we find 
\be 
H_4 - 1 = -  {i  \De_V \De_W  e^{{1 \ov 2} (t_3 + t_4 - t_1 - t_2)} \ov   \sinh { t_{12} \ov 2} \sinh { {t_{34}} \ov 2} } 
+ \cdots  
\ee
which matches the analytic continuation of (4.32) of~\cite{MSY}. 

{Whilst we defer a detailed analysis to future work this discussion can in principle be generalized to higher dimensions to include spatial dependence in each operator insertion. For instance $H_4$ is now replaced by}  
\begin{eqnarray} \
H_4(t_1,x_1,t_2,x_2,t_3,x_3,t_4,x_4) &=& \langle \sP V_1(t_1,x_1) V_2(t_2,x_2) W_1(t_3,x_3) W_3 (t_4,x_4) \rangle  \nonumber \\
&=& \vev{W (t_4,x_4) V (t_2,x_2) W (t_3,x_3) V (t_1,x_1)}  \ .
\end{eqnarray}
In the case of point-wise shift symmetry case the discussion is parallel to the above, and one finds that for  $t_{12}, t_{34} \ll t_{42}$, $x_{12}, x_{34} \ll x_{42}$
\be
H_4-1\sim e^{\lambda (t_{42} - |x_{42}|/v_B )} 
\ee
while the exponentials will again cancel for $G_4$ provided that the effective vertex analogous to \eqref{njm} {is invariant under $\ep_i \to \ep_i + c(x) e^{\pm \lam t}$. Such a condition is satisfied for instance by the effective vertex corresponding to the SYK chains in \cite{Gu}.}

\section{Discussions and conclusions}\label{sec:conc}

Let us first summarize our main findings: with a shift symmetry, the hydrodynamic theory has a mode which grows exponentially 
in time, and exhibits ballistic spreading with a butterfly velocity $v_B$. As a result such behavior appears in OTOCs, 
leading to~\eqref{iem0},~\eqref{iem1}--\eqref{mm0}, while the shift symmetry prevents correlation functions of the energy density and flux, and TOCs of generic operators from having such behavior. 

{A key prediction of this theory is that there should be direct connections between chaos and correlation functions of energy-density and energy flux. In particular we emphasised that within this approach the butterfly velocity is determined by the dispersion relation for a pole in the energy-density two point function according to the relation~\eqref{vb}--\eqref{relationintro}. The fact that $v_B$ is determined by such a pole could provide an explanation for why in many soluble models $v_B$ is comparable to the energy diffusion constant $D_E$~\cite{MB,MB2,MB3,MB4,Davison,Gu}. Furthermore the theory predicts that correlation functions of energy density and flux exhibit the phenomenon of pole-skipping. This phenomenon provides a simpler way than OTOCs to extract the Lyapunov exponent $\lam$ and butterfly velocity $v_B$.   It is of interest to study this pole-skipping phenomenon more generally in holographic systems. }  

While in this paper we considered a theory with only energy conservation we expect the discussion can be straightforwardly  generalized to systems with full energy-momentum conservation, with other conserved quantities, or with additional light modes, since the chaos mode is associated with energy conservation.   
For example, as mentioned earlier, 
strong support for the pole-skipping phenomenon has been observed in a momentum conservation system in~\cite{Saso} 
at exactly the same value of frequency and momentum we indicated. 
%This may be a hint that even the location of the vanishing pole we predicated may still apply to a system with full energy-momentum. 

The hydrodynamic description developed here should also provide new techniques for studying 
chaotic systems, especially those phenomena related to
operator spreading and scrambling, including for example, the spread of entanglement and the traversable wormhole of~\cite{Gao:2016bin}.

There are  still many open questions. One central issue 
is the precise scope of the applicability of the hydrodynamic description. 
A closely related question is why quantum many-body chaos should have anything to do with the hydrodynamic mode $\sigma(t)$ associated to energy conservation? Currently the hints for a hydrodynamic picture which we mentioned in the Introduction appear to largely involve examples with maximal chaos, such as SYK at strong coupling and holographic gravity examples. 
{Further analyses of the 
%In Sec.~\ref{sec:max} we presented several evidences that the analysis of 
TOCs and OTOCs in Sec.~\ref{sec:corre} find further parallels with those of known maximally chaotic systems, which we will elaborate elsewhere. In particular, one finds that the leading exponential growth in the expectation value of a commutator square in fact vanishes due to destructive interference, which appears to resonate well with a recent proposal regarding maximal chaos from~\cite{Kitaev,kitaIAS}.} 
If indeed it turns out that maximally chaotic systems are distinguished from general chaotic systems by having a 
hydrodynamic formulation in terms of $\sigma(t)$ then this would be a rather interesting physical picture and 
open a new window into the physical nature of quantum many-body chaotic behavior. 

It is tempting to further speculate that for a non-maximally chaotic system, proposal A and B of the Introduction
may still apply, i.e. one should still be able to formulate the Lyapunov behavior and butterfly spreading using an effective 
chaotic mode as, for example, in~\cite{Aleiner}. In holographic systems, the Lyapunov exponent deviates from the maximal value when stringy mode exchanges are included~\cite{Shenker2} and it may happen that the net effect of summing over an infinite number of stringy modes can be captured by a single mode as in Regge physics.  Now such a chaos mode may not be 
fully captured by energy conservation. Nevertheless, for a near-maximally chaotic system it appears reasonable that the mode may still have significant overlap with the hydrodynamic mode for energy conservation, and thus there might still be remnant of the  
pole-skipping phenomenon, and there might be some structure in the commutator square as advocated in~\cite{Kitaev,kitaIAS}.

%Whilst the detailed analysis of OTOCs in Section \ref{sec:corre} appears to only apply to maximally chaotic systems, we expect that most of the considerations of this paper should also hold more generally. For instance it should be possible to describe non-maximally chaotic systems in terms of an exponentially growing mode in an effective theory with a shift symmetry. However the precise nature of this exponential mode is currently unclear, and such a theory would require a more general approach to calculating OTOCs than we presented in our discussion in Section~\ref{sec:corre}.

%\textcolor{blue}{Delete this paragraph?}It would be interesting find more support from examples which are not maximally chaotic. So far our description appears to be compatible with a few known such examples including 
%corrections to SYK~\cite{MS,Gu}, stringy corrections in AdS~\cite{Shenker2}, and studies using kinetic theory~\cite{Aleiner,Swingle}.

Another important issue is the  physical origin and nature of the shift symmetry. In the special example of~\eqref{nd3} which provides an effective theory for SYK, the shift symmetry is a subgroup of the $SL(2,R)$ which 
is  a ``global'' gauge symmetry~\cite{MSY}. 
%has been argued in~\cite{MSY} to be a ``global'' gauge symmetry. 
It appears sensible that the shift symmetry should also be  a ``global'' gauge symmetry
in the sense that configurations related by such a shift are considered to be physically equivalent, and thus not integrated over in the path integrals. An interesting question is whether  we can identify the counterpart 
of the shift symmetry on the gravity side in holographic systems. We do not have a full answer, but it 
is tempting to identify this symmetry as a horizon boost, given that on the gravity side the hydrodynamic cloud is essentially 
built up through large relative boosts between stationary observers near a horizon and at the infinity. See also Fig.~\ref{fig:membrane}. One can also show that the exponential solution generated by the shift symmetry precisely matches with 
a shock wave solution on the gravity side, which we will discuss elsewhere. 

%In Appendix~\ref{app:d}  we offer some indirect support for this identification. 

\vspace{0.2in}   \centerline{\bf{Acknowledgements}} \vspace{0.2in}
We would like to thank Koushik Balasubramanian, Zhen Bi, Richard Davison, Aristomenis Donos, Ping Gao, Paolo Glorioso, Saso Grozdanov, Kristan Jensen, Vedika Khemani,  Andy Lucas, Juan Maldacena, Xiaoliang Qi, Srivatsan Rajagopal, Paul Romatschke, Moshe Rozali, Douglas Stanford, Herman Verlinde, Ashvin Vishwanath, and Zhenbin Yang for discussions. HL would in particular like to thank Juan Maldacena for various conversations on the SYK model and AdS$_2$ 
during the summer of 2017, and various correspondences. 
This work is supported by the Office of High Energy Physics of U.S. Department of Energy under grant Contract Number  DE-SC0012567.

\appendix

%\section{Review of formulation of hydrodynamic action} 

%Let us first quickly review the story with full energy-momentum conservation.  

\section{Quadratic action to all derivative orders in the presence of background fields} \label{app:a} 

In this Appendix we discuss the full hydrodynamic action $I_{\mathrm{hydro}}$ to quadratic order in deviations from thermal equilibrium in the presence of background fields. Turning on background fields is needed in determining the explicit forms of $\sE_r, \sJ_r^i, \sE_a, \sJ^i_a$ which are in turn needed to compute various correlation functions of the energy density and energy flux.

Consider infinitesimal deviations from~\eqref{bac} 
\be 
e_r = \ha (e_1 + e_2) = 1 + \fre,  \quad e_a = e_1 - e_2 , \quad w_i = \ha (w_{1i} + w_{2i}), \quad w_{ai} = w_{1i} - w_{2i} \ .
\ee
We will treat the above quantities and deviations~\eqref{uj} of the dynamical fields  from equilibrium 
to be of the same order, and expand the action to quadratic orders 
in these variables.
%and taking $\fre, e_a, w_i , w_{ai}$ and $\ep$ and $\ep_a $ to be small, 
Various variables introduced in Sec.~\ref{sec:eng} can now be written as 
\bega \label{o6}
E_a = e_a - \p_t \ep_a , %- \fre e_a - \ep_a \p_t  \fre + \cdots ,
 \quad \fs_i = \p_i \fre + \p_t  w_i , \qquad
\ft_{ij} = \p_i w_j - \p_j w_i ,
 \\
V_{ai} = w_{ai}  + \p_i \ep_a + \cdots , % \de (w_{ai}  - \p_i \ep_a)
%+ \p_t w_i  \ep_a -  w_i \p_t \ep_a
\qquad  \b =  \b_0 + \de \b , \qquad \de \b =  \b_0(\fre  - \p_t \ep) \ .
% (1 + \fre  - \p_t \ep)
\label{o66}
\end{gather} 
Note that $\fs_i$ and $\ft_{ij}$ depend only on external sources. 

At linear order in these variables we simply find 
\be 
S_1 = a_0 \int d^d x \, e_a  
\ee
with $a_0$ interpreted as the constant part of the energy density. At quadratic order we find 
\be \label{a11}
S_2 = \int d^d x \le(- H E_a + G_i V_{ai} + {i \ov 2} g_1 E_a^2 + i \hat g_2 E_a \p_i V_{ai} + {i \ov 2} (g_3 \p_i \p_j + g_4 \de_{ij}) V_{ai} V_{aj}  \ri)
\ee
where $H$ and $G_i$ can be written more explicitly as 
\be 
H =  f_1 \de \b + f_2 \p_i \fs_i  , \quad G_i =  h_1 \p_i  \de \b + (h_2 \p_j \p_i + h_3 \de_{ij})  \fs_j
+ h_4 (\p_j \de_{ki} - \p_k \de_{ji}) \ft_{jk} \ .
\ee
In writing down~\eqref{a11} we have imposed~\eqref{CC1}--\eqref{CC2}.  
$f_{1,2}, h_{1,2,3},g_{1,2,3,4}$ should be understood as scalar operators constructed from $\p_0$ and $\p_i^2$
acting on the fields immediately behind them. Note $h_4$ is a function of $\p_i^2$ only due to~\eqref{olm}. 
Note that by definition 
\be 
g_1 = g_1^*, \qquad g_3 = g_3^*, \qquad g_4 = g_4^* 
\ee
where $g_1^*$ denotes the operator obtained from $g_1$ by integration by parts. 

At linearized level the dynamical KMS transformations~\eqref{kms} have the form 
\be \label{n09}
\tilde \chi_r (- t) =e^{i (\th - {\b_0 \ov 2}) \p_t}  \le(L \chi_r (t) + {L_a \ov 2} \chi_a (t) \ri), \quad
\tilde \chi_a (-t) = e^{i (\th - {\b_0 \ov 2}) \p_t}  \le(  L \chi_a (t) + 2 L_a \chi_r (t) \ri)
\ee
with
\be 
L = \cosh {i \b_0 \p_t \ov 2} ,
%\ha \le(1 + e^{- i \b_0 \p_t} \ri),
 \qquad L_a =  %{1 \ov 2} \le(1 - e^{- i \b_0 \p_t} \ri)
 \sinh {i \b_0 \p_t \ov 2}
\ee
where $(\chi_r, \chi_a)  = \{(\de E_r, E_a), (V_{ri}, V_{ai})\}$ (recall $\de \b = \b_0 \de E_r$). 
Requiring the action~\eqref{a11} to be invariant under these transformations  we find 
\be \label{wo}
\hat g_2 = g_2 \p_t , \qquad g_2 = g_2^*, \qquad  \b_0 h_1 = - f_2 \p_t -  h_2 \p_i^2 - h_3 
\ee 
and 
\bega \label{w1}
\b_0 (f_1 - f_1^*)= - 2 i \tanh{i \b_0 \p_t \ov 2} M_1 ,  \quad
 \b_0 (h_1 + h_1^*) \p_t  =- 2 i \tanh{i \b_0 \p_t \ov 2} M_2, \\
f_2 - f_2^* =  2i  \tanh {i \b_0 \p_t \ov 2} g_2 , \quad
(h_2 + h_2^*) \p_t  = - 2 i \tanh{i \b_0 \p_t \ov 2} g_3 , \\
(h_3 + h_3^*) \p_t = - 2 i \tanh{i \b_0 \p_t \ov 2}   g_4  \ .
\label{w3}
\end{gather}

From~\eqref{defej}, the symmetric and antisymmetric parts of the energy density and flux can be obtained 
at linearized level as 
 \be 
 \sE_r (t,x^i) = - {\de I_{\rm hydro} \ov \de e_a (t, x^i)}  , \qquad
 \sJ_r^i (t,x^i) = {\de I_{\rm hydro} \ov \de w_{ai} (t, x^i)}
 \ee
and similarly for $\sE_a, \sJ^i_a$ with $e_a, w_{ai}$ in the above equations 
replaced by $\fre, w_i$. We thus find 
\bega \label{uu1}
\sE_r =H - i g_1 E_a  + i  g_2 \p_t  \p_i V_{ai}  , \qquad
 \sE_a =  (\b_0 f_1^* +  f_2^* \p_i^2) E_a + f_2^* \p_t  \p_i V_{ai}
\\ % \b_0 L^* (\p_t \ep_a) -  \b_0 L_i^* (\p_i \ep_a) + \Lam  \ep_a, \\
\sJ^i_r =  G_i  - i  g_2 \p_t \p_i E_a  + i (g_3 \p_i \p_j + g_4 \de_{ij}) V_{aj} , \\ 
\sJ^i_a =  f_2^* \p_t \p_i E_a - (h_2^* \p_i \p_j + h_3^* \de_{ij}) \p_t V_{aj} - 2 h_4^* \p_j w_{aij}  \ 
\label{uu3}
\end{gather}
with $w_{aij} = \p_i w_{aj} - \p_j w_{ai}$. From the above expressions and using the definitions of~\eqref{hj1} and~\eqref{hj2} 
one can find that 
\bega \label{uv1}
\sG_R^{\sE \sE} (x) =  \b_0^2  f_1 h_1 \p_t \p_i^2 G_R -  f_2 \p_i^2, \quad
\sG_R^{\sJ^i\sE} = \sG_R^{\sE \sJ^i} = - \b_0^2  f_1 h_1 \p_t^2 \p_i G_R +  f_2 \p_i \p_t ,   \\ 
\sG_R^{\sJ^i\sJ^j} = - \b_0^2   h_1^2 \p_t^2 \p_j  \p_i G_R  + \tilde h_2 \p_i \p_j + \tilde h_3  \de_{ij} \\
\sG_S^{\sE \sE} (x) = - \b_0^2 f_1 f_1^* \p_t^2 G_S + \b_0 f_1 M_1 \p_t^2 G_R 
+ \b_0 M_1 f_1^* \p_t^2 G_A  + g_1 \\
\sG_S^{\sJ^i \sJ^j} (x) =  \b_0^2 h_1 h_1^* \p_t^2  \p_i \p_j G_S + \b_0 h_1 M_2 \p_t \p_i \p_j G_R 
- \b_0 h_1^* M_2 \p_t \p_i \p_j G_A  + g_3 \p_i \p_j + g_4 \de_{ij} \\
\label{un1}
\sG_S^{\sE \sJ^i} (x) =  \b_0^2 f_1 h_1^* \p_t^2 \p_i G_S + \b_0 f_1 M_2 \p_t \p_i G_R 
- \b_0 M_1 h_1^* \p_t^2 \p_i G_A + g_2 \p_t \p_i  \\
\sG_S^{\sJ^i \sE} (x) = - \b_0^2 f_1^* h_1 \p_t^2 \p_i G_S + \b_0 M_1 h_1 \p_t^2 \p_i G_R + \b_0 f_1^* M_2 \p_t \p_i G_A 
+ g_2 \p_t \p_i  \ 
\label{un}
\end{gather} 
with $\tilde h_2 = h_2 \p_t - 2 h_4,  \tilde h_3 = h_3 \p_t + 2 h_4 \p_i^2$. Note that the terms in the above which do not involve $G_R, G_A, G_S$ are ``contact'' terms. One can check from the above expressions the Onsager relation 
\be 
\sG_S^{\sE \sJ^i} = \sG_S^{\sJ^i \sE} 
\ee
and the fluctuation-dissipation relations 
\be 
\sG_R - \sG_A = 2 i \tanh{i \b_0 \p_t \ov 2} \sG_S 
\ee
for all components.

\section{Equivalence to the Schwarzian action} \label{app:b}

In this Appendix we show that 
the Lagrangian~\eqref{actn} with $H$ given by~\eqref{nd3}, which we copy here for convenience,  
\bega \label{nk}
\sL_{\rm hydro} = - H \p_t X_a + {i \ov 2}  M_1 (\p_t X_a)^2 + O(a^3), \\%\qquad H = H (\b, \p_t \b, \p_t^2 \b , \cdots) \ 
H= - a_2 \Sch (u, t) 
=a_2 \le( {\lam^2 \b_0^2 \ov 2 \b^2} - {\b'^2 \ov 2 \b^2} + {\b'' \ov \b}  \ri) , \quad \b = {\b_0 \ov \p_t \sig}
\end{gather}
can be factorized into two copies of the Schwazian action. 
First note that with $H$ given by the above expression it can be checked that dynamical KMS symmetry~\eqref{kms} requires 
the corresponding $M_1=0$. Following the discussion of~\cite{GL} to construct an entropy, one finds that 
$M_1$ controls the entropy production. Its absence thus implies that 
this theory is non-dissipative, which can also be heuristically deduced from the absence of terms with odd time derivatives in $H$. 

Now writing\footnote{Recall that $X_1 (\sig_1 (t)) = t, X_2 (\sig_2 (t)) = t$.}
\be 
\sig (t) = \ha (\sig_1 (t) + \sig_2 (t)) , \qquad X_a = - {\sig_a \ov \p_t \sig} , \qquad \sig_a = \sig_1 - \sig_2
\ee
we then find that up to total derivatives 
\be 
\sL_{\rm hydro} = \sL [\sig_1] - \sL [\sig_2] + O(\sig^3_a)
\ee
with
\be 
\sL [\sig_1] = - a_2 {\rm Sch} \le(e^{-\lam \sig_1}, t \ri) = a_2 \le({\lam^2 \ov 2} \sig'^2_1 - \Sch (\sig_1, t) \ri) =
a_2 \le({\lam^2 \ov 2} \sig'^2_1 - {\sig'''_1\ov \sig'_1} + {3 \ov 2} {\sig''^2_1 \ov \sig'^2_1}  \ri) \ .
\ee
Note that in the classical limit $\hbar \to 0$~\cite{CGL1}, $O(\sig_a^3)$ terms vanish in both factorized Schwarzian and~\eqref{nk}, so the two theories are completely equivalent. They are also completely equivalent at quadratic order away from equilibrium. 

\section{Wightman Green functions} \label{app:c}

In Sec.~\ref{sec:coeft} we obtained the exponentially growing part of the retarded Green's function $G_R$ for $\sig$. In this Appendix we use the fluctuation-dissipation relation~\eqref{fdt2} to obtain the exponential parts of 
$G_+$ and $G_-$ (defined in~\eqref{gpm}) whose expressions are needed in Sec.~\ref{sec:corre}.  

For this purpose let us write  
\be 
G_R (t)= i \th (t) \De (t), 
\quad G_A =  G_R (-t) = -i \th (-t) \De (t) , \quad \De (t) =- \De (-t) \  .
\ee
then the fluctuation-dissipation relation~\eqref{fdt2} can be written as 
\be   \label{sfdt}
\De (t) = \le(e^{i \b_0 \p_t} -1 \ri) G_- (t) \ 
\ee
from which we can determine the exponential part of $G_- (t)$ from $G_R$. 
Writing $G_R$ as 
\be \label{imm1}
G_R =  \th (t) c e^{\lam t} + \cdots 
\ee
then from~\eqref{sfdt} we find
\be \label{nn1}
G_- (t)  =  \bca  - {i c \ov e^{i \lam \b_0} -1} e^{\lam t} + \cdots & t > 0 \cr
    {i c \ov e^{- i \lam \b_0} -1} e^{- \lam t}  + \cdots  & t < 0
    \eca \quad \Lra \quad
G_+ (t)  =  \bca   {i c \ov e^{- i \lam \b_0} -1} e^{\lam t} + \cdots & t > 0 \cr
    - {i c \ov e^{ i \lam \b_0} -1} e^{ -\lam t}  + \cdots & t < 0
    \eca  \ .    
    \ee 
The above expressions become singular for the maximally chaotic case, with $\lam = {2 \pi \ov \b_0}$. In this case we find that 
the exponential parts are 
\be \label{nn2}
G_- (t)  =  \bca - {c \ov \b_0} t e^{2 \pi  t \ov \b_0}  + a  e^{2 \pi  t \ov \b_0}   & t > 0 \cr
                  {c \ov \b_0} t e^{-{2 \pi  t \ov \b_0}} +  b e^{-{2 \pi  t \ov \b_0}}   & t < 0 
                 \eca, \quad 
                 G_+ (t)  =  \bca - {c \ov \b_0} t e^{2 \pi  t \ov \b_0}  + b e^{2 \pi  t \ov \b_0}    & t > 0 \cr
                  {c \ov \b_0} t e^{-{2 \pi  t \ov \b_0}}  + a  e^{-{2 \pi  t \ov \b_0}}  & t < 0 
                 \eca \
\ee
with $a, b$ undetermined constants satisfying  $c = i (b - a)$. Note the additional $t e^{\lam_{\rm max} t}$ terms. 

To make comparison with explicit Euclidean calculation in the Schwarzian effective action for SYK in~\cite{MSY} 
let us consider the full correlation functions (not just the exponential parts) for the example~\eqref{nd3} for maximal chaos $\lam = \lam_{\rm max}$.
For notational simplicity we will set $\b_0 = 2 \pi$ so that $\lam =1$ and $a_2 =1$ in~\eqref{nd3}. Apply~\eqref{nem} to the example we find that  
\be 
G_R  =\th (t) (t - \sinh t)
\ee
and from~\eqref{fdt2} we find that 
\be \label{gs2} 
G_S (t) =   - {1 \ov 2 \pi} \le[{t^2 - \pi^2 \ov 2}  - t \sinh t + a + b \cosh t \ri] \ .
\ee
and 
\be \label{gpp}
G_+(t)  = - {1 \ov 2 \pi} \le[{t^2 - \pi^2 \ov 2}  - t \sinh t + a + b \cosh t 
+ i \pi t - i \pi \sinh t \ri] \ .
\ee
where $a,b$ are some undetermined integration constants. Note that~\eqref{gpp} precisely agrees with the Lorentzian analytic continuation of (4.28) of~\cite{MSY}.

\end{document}